\documentclass[12pt]{article}

\usepackage[T1]{fontenc}
\usepackage{lmodern} 
\usepackage[margin=1in]{geometry}
\usepackage{microtype}
\usepackage{amsmath,amssymb,amsthm,mathtools,bm}

\usepackage{booktabs,array,longtable}
\usepackage{graphicx}
\usepackage{lscape}
\usepackage{float}
\usepackage{multirow}
\usepackage{enumitem}
\usepackage[authoryear,round]{natbib}
\usepackage{xcolor}
\usepackage{tikz}
\usepackage[
  colorlinks=true,
  linkcolor=blue,
  citecolor=blue,
  urlcolor=blue
]{hyperref}
\usepackage{url}
\allowdisplaybreaks
\usepackage{indentfirst}
\usepackage{setspace}
\newtheorem{lemma}{Lemma}
\newtheorem{theorem}{Theorem}
\newtheorem{proposition}{Proposition}
\newtheorem{assumption}{Assumption}
\newtheorem{remark}{Remark}

\newcommand{\E}{\mathbb E}
\newcommand{\Pp}{P}
\newcommand{\Pn}{\mathbb P_n}
\newcommand{\Prb}{\mathbb P}
\newcommand{\ind}{\mathbb I}
\newcommand{\cb}{\mathrm{cb}}

\newcommand{\dr}{\mathrm{dr}}
\newcommand{\opt}{\mathrm{opt}}

\DeclareMathOperator*{\argmax}{arg\,max}
\DeclareMathOperator*{\argmin}{arg\,min}
\newcommand{\subX}{\scriptscriptstyle\mathrm{X}}
\newcommand{\subZ}{\scriptscriptstyle\mathrm{Z}}
\newcommand{\subL}{\scriptscriptstyle\mathrm{L}}
\newcommand{\subC}{\scriptscriptstyle\mathrm{C}}
\newcommand{\subI}{\scriptscriptstyle\mathrm{I}}
\newcommand{\true}{\circ}

\newcommand{\indep}{\perp\!\!\!\perp}

\title{\bf\LARGE Covariate Balancing Value Estimation for Optimal Individualized Treatment Rules}
\author{\bf\small Yue Zhang$^1$, Shanshan Luo$^{2,\ast}$, Zhi Geng$^{1,2}$, and Yangbo He$^{1,\ast}$\\
\small $^1$School of Mathematical Sciences, Peking University, Beijing, China \\
\small $^2$School of Mathematics and Statistics, Beijing Technology and Business University, Beijing, China}
\date{}

\begin{document}
\maketitle
\begingroup
\renewcommand{\thefootnote}{\ensuremath{\ast}}
\footnotetext{Corresponding authors: Shanshan Luo (\href{mailto:shanshanluo@btbu.edu.cn}{shanshanluo@btbu.edu.cn}) and Yangbo He (\href{mailto:heyb@math.pku.edu.cn}{heyb@math.pku.edu.cn}).}
\endgroup

\begin{abstract}
Learning an optimal individualized treatment rule depends on reliable value comparisons across the candidate class. Standard doubly robust estimators are consistent when either the propensity score or outcome regression model is correctly specified, but they do not directly control the remaining bias in value estimation when both models are misspecified. 
In this paper, we propose a covariate balancing doubly robust estimator that combines propensity score estimation based on covariate balancing with an outcome regression component selected using an empirical variance criterion based on the influence function.
Using prespecified covariate functions, the balancing procedure induces an effective balancing space to which the weighted propensity score error is orthogonal.
Consequently, the proposed value estimator is consistent if either the propensity score model is correct or the rule-relevant outcome regression error lies in this space. The latter condition does not require a correctly specified outcome regression model and can hold even when both working models are misspecified.
Under correct propensity score specification, the estimator has the smallest asymptotic variance within the proposed covariate balancing doubly robust class.
Simulations evaluate value estimation in finite samples and the regret of learned rules, and an application to a leukemia dataset illustrates the proposed method in practice.
\end{abstract}


\section{Introduction}

Precision medicine seeks to tailor treatment decisions to individual patient characteristics, a goal commonly formalized through individualized treatment rules that maximize expected clinical outcomes \citep{manski2004statistical}. Selecting an optimal rule is ultimately a problem of comparing the values of candidate rules. These values are counterfactual because each patient is observed under only the treatment actually received. Bias can systematically rank candidate rules incorrectly, while inefficient estimates can make their ranking unstable across samples. Existing methods approach rule learning either indirectly, by deriving a rule from a fitted model of how treatment affects outcomes \citep{murphy2003optimal,qian2011performance}, or directly, by optimizing an estimated value over a prespecified rule class \citep{zhao_estimating_2012,Zhang2012,zhou2017residual,laber2015tree,athey2021policy}. 
More generally, value estimation provides a common basis for comparing and evaluating rules learned by different approaches.
Therefore both robustness and efficiency are required for reliable value estimation. 

Augmented inverse probability weighted estimators provide an appealing foundation for value estimation by combining propensity score and outcome regression models to guard against model misspecification. For a given treatment rule, the standard doubly robust estimator is consistent if either nuisance model is correctly specified \citep{Robins1994JASA,Zhang2012}. A common implementation estimates the propensity score by maximum likelihood and fits the outcome regression by least squares \citep{Zhang2012}, while \citet{pan2021improved} selects the outcome parameter to reduce asymptotic variance under a correct propensity score model. Neither strategy, however, directly controls the remaining value bias when both working models are misspecified \citep{kang2007demystifying}.  \citet{vermeulen_bias-reduced_2015} propose choosing nuisance parameters by locally minimizing the squared first-order asymptotic bias, yielding local bias reduction even when both nuisance models are misspecified.

Covariate balancing offers another route to controlling nuisance misspecification bias. It estimates the propensity score by choosing weights that balance the treatment groups with respect to prespecified covariate functions \citep{imai2014covariate,Fan2021OptimalCB}. In the policy learning literature, \citet{li2025robust} combine covariate balancing with matching to estimate a contrast value, obtaining consistency when either the propensity score model or the matching component is correct. 
The key observation motivating our approach is that the residual value bias of the standard doubly robust estimator can be expressed as an inner product between a weighted propensity score error and a rule-relevant outcome regression error. When the prespecified functions effectively balance this outcome error, a requirement that we formalize as a span condition in this paper, the two errors are orthogonal and the bias vanishes.
We therefore incorporate covariate balancing into a doubly robust value estimator and characterize the effective balancing space generated by potentially overidentified balancing equations. To improve efficiency in addition to robustness, we further select the outcome regression component using an empirical variance criterion based on the influence function.

Our contributions are threefold. 
First, we develop a covariate balancing doubly robust estimator whose outcome parameter minimizes an empirical variance criterion based on the influence function. Second, we establish consistency under either a correct propensity score model or a span condition on the rule-relevant outcome regression error, which can hold under joint nuisance-model misspecification. Third, we establish asymptotic variance optimality under correct propensity score specification, asymptotic normality when either the propensity score model is correctly specified or the span condition holds, and other desirable asymptotic properties. 
Simulations examine value estimation in finite samples and the regret of learned rules across different scenarios, with the clearest gains under joint misspecification and broadly comparable performance in the remaining scenarios.
An application to leukemia data demonstrates the implementation of the proposed method in an observational clinical setting.

The remainder of the paper is organized as follows. Section~\ref{sec:setup} introduces the setup and motivates covariate balancing for value estimation. Section~\ref{sec:cb-dr-var-opt} presents the proposed covariate balancing doubly robust estimator, and Section~\ref{sec:balancing-robustness} develops its robustness and efficiency properties. Section~\ref{sec:simulation} and Section~\ref{sec:application} report the numerical simulations and leukemia data application, respectively, and Section~\ref{sec:discussion} concludes. 
The Supplementary Material extends the analysis to auxiliary outcome predictors, establishes an efficiency comparison result, reports additional simulations and sensitivity analyses, and provides regularity conditions and proofs.

\section{Framework}
\label{sec:setup}
\subsection{Setup}

Consider an experimental or observational study involving $n$ participants, each independently and identically sampled from a target population. For one participant, write $O=(Y,A,X)$ for the observed data, where $X\in\mathcal X$ is a vector of pretreatment covariates, $A\in\{0,1\}$ is a binary treatment, and $Y$ is a continuous outcome, with larger values being preferable.

An individualized treatment rule is a function $d: \mathcal{X} \rightarrow \{ 0, 1 \}$ that assigns treatment $d(x)$ to an individual with covariates $X=x$. 
Let $Y(a)$ denote the potential outcome under treatment $a\in\{0,1\}$ \citep{Rubin1974}. 
The performance of rule $d$ is measured by the marginal mean outcome $V(d) = \E\{Y(d)\},$ also known as the value function, where $Y(d) = Y(1) d(X) + Y(0) \{ 1 - d(X) \}$ is the potential outcome under rule $d$.
Our goal is to develop an optimal treatment rule within a prespecified class $\mathcal{D}$, namely, $d^{\opt} \in \argmax_{d\in\mathcal D} V(d)$.

Let \(e_a^{\true}(X)=\Prb(A=a\mid X)\) denote the true propensity score \citep{ROSENBAUM1983Bka}, and let \(m_a^{\true}(X)=\E\{Y(a)\mid X\}\) denote the true conditional potential outcome mean under treatment \(a\).
We assume the stable unit treatment value assumption in this paper, so that there is no interference between participants and each treatment level corresponds to a well-defined intervention \citep{rubin1980randomization}.
The following assumption is invoked throughout to ensure identifiability of the value function.

\begin{assumption}\label{assump:regular}
(i) $Y = AY(1) + (1-A)Y(0)$; (ii) $\varepsilon < e_a^{\true} (X) < 1 - \varepsilon$ almost surely for $a \in \{0,1\}$ and some $\varepsilon > 0$; (iii) $A \indep \{Y(1),Y(0)\} \mid X$.
\end{assumption}

Under Assumption~\ref{assump:regular}, the value function $V(d)$ can be identified by the following equivalent representations \citep{qian2011performance}:
\begin{equation*}
V(d) = \E \left\{m_d^{\true}(X)\right\} = \E \left[\frac{\ind\{A=d(X)\}Y}{e_d^{\true}(X)}\right],
\end{equation*}
where \(e_d^{\true}(X)=d(X)e_1^{\true}(X)+\{1-d(X)\}e_0^{\true}(X)\) and \(m_d^{\true}(X)=d(X)m_1^{\true}(X)+\{1-d(X)\}m_0^{\true}(X)\) are the propensity score and conditional outcome mean induced by the given rule $d$, respectively. 
In observational studies, both nuisance functions are generally unknown and must be estimated from observed data. Let \(e_a(X;\alpha)\) and \(m_a(X;\beta)\) denote working models for the propensity score and outcome regression, respectively, where $\alpha$ and $\beta$ are finite-dimensional parameters. The augmented inverse probability weighted estimator of \(V(d)\) is
\begin{equation}\label{eq:value-dr-emp}
\widehat V^{\dr}(d;\widehat\alpha,\widehat\beta)
=\Pn\left[
m_d(X;\widehat\beta)
+\frac{\ind\{A=d(X)\}}{e_d(X;\widehat\alpha)}
\left\{Y-m_d(X;\widehat\beta)\right\}
\right],
\end{equation}
where \(\Pn f=n^{-1}\sum_{i=1}^n f(O_i)\) for any integrable function \(f\).
The estimator \eqref{eq:value-dr-emp} is doubly robust, as indicated by the superscript \(\dr\), because it is consistent if either nuisance model is correctly specified \citep{Robins1994JASA}. 

\subsection{Motivation}
To reveal the bias structure of \eqref{eq:value-dr-emp}, suppose that \(\widehat\alpha\overset{p}{\to}\alpha^*\) and \(\widehat\beta\overset{p}{\to}\beta^*\). Let \(V^*(d)\) denote the probability limit of \(\widehat V^{\dr}(d;\widehat\alpha,\widehat\beta)\). Then the asymptotic bias of the estimator \eqref{eq:value-dr-emp} is 
\begin{equation}\label{eq:dr-value-bias-product}
V^*(d)-V(d) =\E\left[\frac{e_d^{\true}(X)-e_d(X;\alpha^*)}{e_d(X;\alpha^*)}\{m_d^{\true}(X)-m_d(X;\beta^*)\}\right].
\end{equation}
The representation \eqref{eq:dr-value-bias-product} recovers the usual doubly robust property of \eqref{eq:value-dr-emp}: for \(a\in\{0,1\}\), the bias vanishes if either (i) there exists \(\alpha^{\true}\) such that \(\alpha^*=\alpha^{\true}\) and \(e_a(X;\alpha^{\true})=e_a^{\true}(X)\), or (ii) there exists \(\beta^{\true}\) such that \(\beta^*=\beta^{\true}\) and \(m_a(X;\beta^{\true})=m_a^{\true}(X)\).
The usual doubly robust estimator obtains \(\widehat\alpha\) by maximum likelihood and \(\widehat\beta\) by ordinary least squares in \eqref{eq:value-dr-emp} \citep{Zhang2012}. The improved doubly robust estimator of \citet{pan2021improved} selects the outcome parameter to reduce asymptotic variance while retaining propensity score estimation by maximum likelihood. Neither approach, however, directly targets the remaining bias in \eqref{eq:dr-value-bias-product} when both nuisance models are misspecified, which is determined by the population inner product of two error terms induced by the rule and the nuisance models.

The product structure in \eqref{eq:dr-value-bias-product} shows that eliminating the remaining bias does not require either nuisance error to vanish; it is sufficient for the two errors to be orthogonal. Covariate balancing provides a natural way to induce this orthogonality and can therefore protect value estimation against joint model misspecification. Motivated by this observation, we combine a propensity score estimator based on balancing with an outcome regression component chosen to improve efficiency, yielding the proposed estimator in the next section.

\section{Covariate balancing doubly robust estimation}\label{sec:cb-dr-var-opt}
In this section, we develop an estimation strategy that combines propensity score estimation by covariate balancing with selection of the outcome regression parameter using a variance criterion based on the influence function. We first summarize the proposed procedure below, and then develop its two components in Sections~\ref{sec:cb-ps-est} and \ref{sec:var-opt-outcome}, respectively.

\paragraph{Estimation procedure.}
\begin{enumerate}[label={\it Step \arabic*:}, leftmargin=50pt]
	\item Estimate the propensity score by balancing a prespecified set of covariate functions, yielding the covariate balancing propensity score estimator \(\widehat\alpha^{\cb}\) (Section~\ref{sec:cb-ps-est}).
	\item For each candidate rule \(d\), select an outcome regression parameter, denoted by \(\widehat\beta^{\opt}\), by minimizing an empirical variance criterion based on the influence function within the resulting estimator class (Section~\ref{sec:var-opt-outcome}).
	\item Obtain the proposed \emph{covariate balancing doubly robust estimator} by evaluating \eqref{eq:value-dr-emp} at $(\widehat{\alpha},\widehat{\beta})=(\widehat{\alpha}^{\cb},\widehat{\beta}^{\opt})$, that is,
	\begin{equation}\label{eq:value-cb-emp}
		\widehat V^{\cb}(d)
		=\widehat V^{\dr}(d;\widehat\alpha^{\cb},\widehat\beta^{\opt}).
	\end{equation}
	\item Select an estimated optimal individualized treatment rule in the prespecified rule class \(\mathcal D\) by maximizing the proposed value estimator:
	\[
		\widehat d\in\argmax_{d\in\mathcal D}\widehat V^{\cb}(d).
	\]
\end{enumerate}

We refer to the resulting estimator as covariate balancing doubly robust because its consistency can follow from either a correctly specified propensity score model or the balancing condition developed from covariate balancing in Section~\ref{sec:balancing-space-bias}.

\subsection{Covariate balancing propensity score estimation}\label{sec:cb-ps-est}

We next incorporate covariate balancing into the propensity score component of the doubly robust estimator \eqref{eq:value-dr-emp}. Following \citet{imai2014covariate}, let $h(X;\alpha)$ be a finite-dimensional balancing basis that might depend on the propensity score model, and define the following moment function:
\begin{equation}\label{eq:cbps-eq}
	\psi(A,X;\alpha)=\frac{A h(X;\alpha)}{e_1(X;\alpha)}-\frac{(1-A)h(X;\alpha)}{1 - e_1(X;\alpha)}.
\end{equation}
The empirical average of $\psi$ measures the difference between the inverse probability weighted totals of the balancing functions $h$ in the treated and control groups.
In practice, \(h\) may be prespecified using low-order functions of the baseline covariates, such as \(h(X)=\{1,X^\top,(X^2)^\top\}^\top\). A richer basis may improve bias control but can also increase numerical instability, as illustrated by the simulations in Section~\ref{sec:simulation} and the Supplementary Material.

Let $\Psi(\alpha)=\E\{\psi(A,X;\alpha)\}$.
The covariate balancing propensity score estimator is obtained by minimizing the sample GMM criterion \citep{Hansen1982}:
\begin{equation}\label{eq:cb-estimating-equation}
\widehat\alpha^{\cb}\in\argmin_{\alpha}\left\|\widehat\Psi_n(\alpha)\right\|_2^2,
\end{equation}
where $\|\cdot\|_2$ denotes the Euclidean norm and $\widehat\Psi_n(\alpha)=\Pn\{\psi(A,X;\alpha)\}$ is the sample analogue of $\Psi(\alpha)$. Throughout, we use the identity weighting matrix in the GMM procedure \eqref{eq:cb-estimating-equation}. 
Thus, \eqref{eq:cb-estimating-equation} directly minimizes covariate imbalance in the prespecified functions. In an exactly identified system, when exact balance is attainable, it reduces to $\widehat\Psi_n(\widehat\alpha^{\cb})=0$.

Let $\alpha^*$ be the unique minimizer of $\|\Psi(\alpha)\|_2^2$ and define $G(\alpha)=\partial\Psi(\alpha)/\partial\alpha^\top$. Under correct propensity score specification, we have $\alpha^*=\alpha^{\true}$ and $\Psi(\alpha^{\true})=0$. Set $G^{\true}=G(\alpha^{\true})$ and
$K=\{(G^{\true})^\top G^{\true}\}^{-1}(G^{\true})^\top$.  The expansion of the estimator in \eqref{eq:cb-estimating-equation} 
\begin{equation}\label{eq:alpha-if}
\sqrt n\left(\widehat\alpha^{\cb}-\alpha^{\true}\right) =-K\frac{1}{\sqrt n}\sum_{i=1}^n \psi(A_i,X_i;\alpha^{\true})+o_p(1)
\end{equation}
holds under the regularity conditions in Assumption~\ref{assump:app-cb-regularity} of the Supplementary Material, which is maintained throughout.

\subsection{Variance-optimized outcome regression estimation}\label{sec:var-opt-outcome}
Substituting the covariate balancing propensity score estimator \(\widehat\alpha^{\cb}\) derived in Section~\ref{sec:cb-ps-est} for \(\widehat\alpha\) in \eqref{eq:value-dr-emp} yields the following class of covariate balancing doubly robust estimators of $V(d)$:
\begin{equation}\label{eq:cb-value-class}
\widehat V^{\cb}(d;\beta)
=\Pn\left[
m_d(X;\beta)
+\frac{\ind\{A=d(X)\}}{e_d(X;\widehat\alpha^{\cb})}
\left\{Y-m_d(X;\beta)\right\}
\right].
\end{equation}
We next choose \(\beta\) to minimize the asymptotic variance within the class \eqref{eq:cb-value-class}.
First we establish the first-order behavior of the estimator in \eqref{eq:cb-value-class} when the propensity score working model is correctly specified.
Combining \eqref{eq:alpha-if} with a first-order expansion of \eqref{eq:value-dr-emp} shows that estimating the covariate balancing propensity score parameter contributes an additional term to the influence function, as stated in the following lemma.

\begin{lemma}\label{lem:cb-if}
Under Assumption~\ref{assump:regular}, suppose the propensity score working model is correctly specified, that is, $e_a(X;\alpha^{\true})=e_a^{\true}(X)$ for $a\in\{0,1\}$ and some $\alpha^{\true}$.
Fix a treatment rule \(d\in\mathcal D\).
If $\widehat\beta$ is root-$n$ consistent for a probability limit $\beta^*$, then
\begin{equation}\label{eq:value-if-expansion}
\sqrt n\left\{\widehat V^{\cb}(d;\widehat\beta)-V(d)\right\} =\frac{1}{\sqrt n}\sum_{i=1}^n \varphi^{\cb}(O_i;\alpha^{\true},\beta^*)+o_p(1).
\end{equation}
Here $\varphi^{\cb}$ is the influence function for the estimator in \eqref{eq:cb-value-class}, defined by
\begin{equation}\label{eq:cb-value-if}
\varphi^{\cb}(O;\alpha^{\true},\beta^*) = \varphi^{\dr}(O;\alpha^{\true},\beta^*) +\Gamma(\alpha^{\true},\beta^*)K \psi(A,X;\alpha^{\true}),
\end{equation}
with \(\Gamma(\alpha,\beta)=-\E\{\partial \varphi^{\dr}(O;\alpha,\beta) / \partial \alpha^\top\}\) and
\begin{equation}\label{eq:dr-if}
\varphi^{\dr}(O;\alpha,\beta)=m_d(X;\beta)+\frac{\ind\{A=d(X)\}}{e_d(X;\alpha)}\left\{Y-m_d(X;\beta)\right\}-V(d).
\end{equation}
\end{lemma}

 Lemma~\ref{lem:cb-if} does not require correct specification of the outcome regression model, implying that \(\beta^*\) may differ from \(\beta^{\true}\). 
Under correct propensity score specification, the influence function \eqref{eq:cb-value-if} does not depend on the first-order sampling variation of \(\widehat\beta\), but only on its probability limit \(\beta^*\). 
If the outcome regression is also correctly specified, so that \(\beta^*=\beta^{\true}\), then \(\Gamma(\alpha^{\true},\beta^{\true})=0\). The second term in \eqref{eq:cb-value-if} therefore vanishes, reducing \(\varphi^{\cb}\) to the standard doubly robust influence function \(\varphi^{\dr}\) in \eqref{eq:dr-if} \citep{Robins1994JASA,tsiatis2007semiparametric}.

It follows from Lemma~\ref{lem:cb-if} that the asymptotic variance of $\widehat V^{\cb}(d)$ is proportional to \(\mathcal S(\alpha^{\true},\beta^*)\), where \(\mathcal S(\alpha^{\true},\beta^*)=\E[\{\varphi^{\cb}(O;\alpha^{\true},\beta^*)\}^2]\).
The optimal outcome parameter minimizes the asymptotic variance within the class \eqref{eq:cb-value-class}  \citep{pan2021improved}, that is, \(\beta^{\opt}\in\argmin_\beta\mathcal S(\alpha^{\true},\beta)\). 
Let $\widehat{\mathcal S}_n(\alpha,\beta)$ denote the empirical version of $\mathcal S(\alpha,\beta)$. 
We then choose the outcome parameter that attains the smallest asymptotic variance within the estimator class \eqref{eq:cb-value-class}:
\[
\widehat\beta^{\opt}\in\argmin_\beta\widehat{\mathcal S}_n(\widehat\alpha^{\cb},\beta).
\]
Finally, setting \(\widehat\beta=\widehat\beta^{\opt}\) in \eqref{eq:cb-value-class} gives the proposed estimator \(\widehat V^{\cb}(d)\) defined in \eqref{eq:value-cb-emp}.

\section{Robustness and efficiency under covariate balancing}
\label{sec:balancing-robustness}

\subsection{Value consistency through covariate balancing orthogonality}
\label{sec:balancing-space-bias}

Now we derive the orthogonality implied by the covariate balancing criterion \eqref{eq:cb-estimating-equation} and use it to establish consistency conditions for the proposed estimator. 
Note that the bias in \eqref{eq:dr-value-bias-product} can then be written as
\begin{equation}\label{eq:dr-value-decomp}
V^*(d)-V(d) = \E\left\{\kappa(X)q_d(X)\right\},
\end{equation}
where $\kappa(X)=\{e_1^{\true}(X)-e_1(X;\alpha^*)\}/[e_1(X;\alpha^*)\{1-e_1(X;\alpha^*)\}]$ is referred to as the weighted propensity score error, and $q_d(X)=d(X)\{1-e_1(X;\alpha^*)\}\{m_1^{\true}(X)-m_1(X;\beta^*)\}-\{1-d(X)\}e_1(X;\alpha^*)\{m_0^{\true}(X)-m_0(X;\beta^*)\}$ is referred to as the rule-relevant outcome regression error. 
The quantity $q_d$ collects the outcome regression error in the treatment arm selected by $d$, weighted by the working propensity score for the opposite arm.


We next show how the balancing criterion controls the bias \eqref{eq:dr-value-decomp}.
Recall that $\Psi(\alpha)=\E\{\psi(A,X;\alpha)\}$, where $\psi$ is the covariate balancing moment function in \eqref{eq:cbps-eq}, and $G(\alpha)=\partial\Psi(\alpha)/\partial\alpha^\top$ as defined in Section~\ref{sec:cb-ps-est}. 
Write $G^*=G(\alpha^*)$. Then the population first-order condition induced by the covariate balancing procedure in \eqref{eq:cb-estimating-equation} is
\begin{equation}\label{eq:cb-population-foc}
(G^*)^\top\Psi(\alpha^*)=0.
\end{equation}
Since \(\Psi(\alpha^*)=\E\{\kappa(X)h(X;\alpha^*)\}\), we consider the following linear span of the balancing functions $h(X;\alpha)$:
\[
\mathcal H_{\rm eff}=\left\{c^\top(G^*)^\top h(X;\alpha^*):c\in\mathbb R^{\dim(\alpha)}\right\}.
\]
We refer to $\mathcal H_{\rm eff}$ as the \emph{effective balancing space}, which consists of the directions in the span of the balancing functions that are enforced by the population first-order condition \eqref{eq:cb-population-foc}.
For every $f\in\mathcal H_{\rm eff}$, Condition~\eqref{eq:cb-population-foc} gives $\E\{\kappa(X)f(X)\}=0$. Thus, the balancing criterion makes $\kappa$ orthogonal to $\mathcal H_{\rm eff}$, and the bias \eqref{eq:dr-value-decomp} vanishes whenever $q_d\in\mathcal H_{\rm eff}$. This yields the following consistency result.

\begin{proposition}
\label{prop:double-robustness}
Suppose Assumption~\ref{assump:regular} holds.
Then $\widehat V^{\cb}(d)\overset{p}{\to} V(d)$ for a fixed rule $d\in\mathcal D$ if either
\begin{enumerate}[label=(\roman*)]
\item the propensity score working model is correctly specified; or
\item the rule-relevant outcome regression error lies in the effective balancing space:
\begin{equation}\label{eq:qd-effective-space}
q_d(X)\in\mathcal H_{\rm eff}.
\end{equation}
\end{enumerate}
\end{proposition}

Proposition~\ref{prop:double-robustness}(i) shows that, under a correct propensity score model, $\kappa=0$ and the bias \eqref{eq:dr-value-decomp} vanishes regardless of the outcome model.
Under part~(ii) of Proposition~\ref{prop:double-robustness}, the population first-order condition \eqref{eq:cb-population-foc} implies that the bias in \eqref{eq:dr-value-decomp} is zero whenever the span condition~\eqref{eq:qd-effective-space} holds. 
This condition can hold even when both working models are misspecified, generalizing the usual double robustness because when the outcome working model is correctly specified while the propensity score model is not, we have \(q_d=0\) if \(\beta^*=\beta^{\true}\) and hence Condition~\eqref{eq:qd-effective-space} holds automatically. Correct specification of the outcome working model alone, however, need not imply \(\beta^*=\beta^{\true}\) for the covariate balancing doubly robust estimator $\widehat V^{\cb}(d)$, because \(\beta^*\) minimizes the variance criterion rather than the least-squares criterion used to fit the outcome regression model. Supplementary Section~\ref{sec:app-or-route} gives a sufficient balancing condition for consistency in this setting. 

Although the span condition~\eqref{eq:qd-effective-space} is generally not directly verifiable because \(q_d\) depends on unknown population quantities, its implied orthogonality yields
\[
\left|V^*(d)-V(d)\right|
\leq \|\kappa\|_{L^2(P)}\,\|q_d-f\|_{L^2(P)},
\]
for any \(f\in\mathcal H_{\rm eff}\), where $\|g\|_{L^2(P)}=[\E\{g(X)^2\}]^{1/2}$. 
Thus, a prespecified basis can still control residual bias when \(q_d\) is well approximated by \(\mathcal H_{\rm eff}\), even if the exact inclusion condition in \eqref{eq:qd-effective-space} does not hold.
The numerical results in Section~\ref{sec:simulation} provide evidence for this bias-reduction property of the proposed estimator in the simulation settings considered.

\subsection{Asymptotic properties}
\label{sec:cb-asymptotics}
We first summarize the asymptotic properties of the proposed value estimator for a prespecified treatment rule and then give a regret bound for the learned rule.

\begin{proposition}\label{prop:asymptotic-properties}
Fix a treatment rule \(d\in\mathcal D\).
Under Assumption~\ref{assump:regular}, the proposed estimator \(\widehat V^{\cb}(d)\) has the following properties:
\begin{enumerate}[label=(\roman*)]
\item If the propensity score working model is correctly specified, \(\widehat V^{\cb}(d)\) attains the smallest asymptotic variance among the estimators in the class \eqref{eq:cb-value-class}.

\item If either condition in Proposition~\ref{prop:double-robustness} holds, then
\[
\sqrt n\left\{\widehat V^{\cb}(d)-V(d)\right\}
\overset{d}{\longrightarrow}N(0,\Lambda),
\]
where the detailed expression for \(\Lambda\) is given in Supplementary Section~\ref{sec:app-proofs}.

\item If both the propensity score and outcome regression working models are correctly specified, then \(\widehat V^{\cb}(d)\) attains the semiparametric efficiency bound.
\end{enumerate}
\end{proposition}


Using a consistent plug-in sandwich estimator of \(\Lambda\) based on Proposition~\ref{prop:asymptotic-properties}(ii) enables Wald inference on \(V(d)\) given a prespecified rule. 
For rule learning, let \(\mathcal D_n\subseteq\mathcal D\) denote a possibly growing candidate class at sample size \(n\), let \(d_n^{\opt}\in\argmax_{d\in\mathcal D_n}V(d)\), and let \(\widehat d_n\in\mathcal D_n\) denote the learned optimal individualized rule over \(\mathcal D_n\). The regret bound separates four sources of error: the complexity of the candidate rule class, nuisance parameter estimation, residual bias in value estimation, and approximate empirical maximization.
Under the regularity conditions in Supplementary Section~\ref{sec:regret}, the regret satisfies
\[
V(d_n^{\opt})-V(\widehat d_n)
=O_p\left(
\sqrt{\frac{v_n}{n}}
+r_{\alpha,n}+r_{\beta,n}+b_n+\rho_n
\right).
\]
Here \(v_n\) measures the complexity of \(\mathcal D_n\), and \((r_{\alpha,n},r_{\beta,n})\) bound the uniform effect of estimating the nuisance parameters. The term \(b_n\) is the largest residual population bias in value estimation over \(\mathcal D_n\), and \(\rho_n\) is the error from approximate empirical maximization. In particular, \(b_n=0\) if either condition in Proposition~\ref{prop:double-robustness} holds.
The precise definitions and conditions are provided in Supplementary Section~\ref{sec:regret}.

\section{Simulations}
\label{sec:simulation}

In this section, we evaluate the performance of the proposed covariate balancing doubly robust estimator in finite samples. The outcome parameter is selected by optimizing a variance criterion based on the influence function; hence, we denote the proposed estimator by \texttt{CB-opt}. We compare it with three benchmarks: the usual doubly robust estimator of \citet{Zhang2012} (\texttt{Usual}), the improved doubly robust estimator of \citet{pan2021improved} (\texttt{Improved}), and the covariate balancing estimator with the outcome parameter estimated by ordinary least squares (\texttt{CB-ols}), which belongs to the estimator class \eqref{eq:cb-value-class}. 
We evaluate the four methods from two aspects: (i) value estimation for a prespecified treatment rule and (ii) regret of the learned optimal treatment rule.

We generate the data as follows. The decision covariates are generated independently as $X_1\sim\mathrm{Unif}(-1,1)$ and $X_2\sim\mathrm{Unif}(-2,2)$. We also generate auxiliary predictors of the outcome, $W_1\sim\mathrm{Unif}(-1,1)$ and $W_2\sim\mathrm{Unif}(-2,2)$, independently of $X$. These predictors affect the outcome but neither treatment assignment nor the class of treatment rules. 

The treatment is generated by $A\mid X\sim\mathrm{Bernoulli}\{e_\star(X)\}$, where $\star\in\{\texttt{C},\texttt{I}\}$. We consider two choices for the true propensity score:
\[
\begin{aligned}
e_{\subC}(X)=\operatorname{expit}\left(2\{1.2X_1-0.2X_2\}\right), ~~
e_{\subI}(X)=\operatorname{expit}\left(-0.5+2\{1.2X_1-0.2X_2\}+\delta|X_1X_2|\right),
\end{aligned}
\]
where $\operatorname{expit}(u)=(1+e^{-u})^{-1}$.
The parameter $\delta$ controls the nonlinear departure from the propensity score working model and $\delta=1$ in this setting. We estimate the propensity score using a logistic working model linear in \(\{1,X_1,X_2\}\). Thus this working model is correctly specified under \(e_{\subC}\) and misspecified under \(e_{\subI}\).

For the outcome, we consider two choices for the true conditional mean given $(X,W)$:
\begin{equation}\label{eq:simulation-outcome-means}
\begin{aligned}
\widetilde m_{a,\subC}^{\true}(X,W)=10\mu_{\mathrm{lin}}(a,X,W), ~~
\widetilde m_{a,\subI}^{\true}(X,W)=10\left\{\mu_{\mathrm{lin}}(a,X,W)+\lambda\mu_{\mathrm{nl}}(a,X,W)\right\},
\end{aligned}
\end{equation}
where
\[
\begin{aligned}
\mu_{\mathrm{lin}}(a,X,W)
&=4X_1-X_2+3W_1-2W_2+(2a-1)(1-2X_1+X_2),\\
\mu_{\mathrm{nl}}(a,X,W)
&=4X_1^2-W_2^2+W_1X_2+5W_2X_1
 +(2a-1)(1-2X_1+X_2)|X_2|.
\end{aligned}
\]
The parameter $\lambda$ controls the nonlinear departure from the outcome regression working model and $\lambda=1$ in this setting. The observed outcome is generated by $Y=\widetilde m_{A,\star}^{\true}(X,W)+\varepsilon$, where $\star\in\{\texttt{C},\texttt{I}\}$ and $\varepsilon\sim N(0,1)$ is independent of $(A,X,W)$. In this section, we use the basis \(\{1,A,X_1,X_2,AX_1,AX_2\}\) to model the marginal mean \(\E(Y\mid A,X)\), which correctly specifies the marginal mean obtained from \(\widetilde m_{a,\subC}^{\true}(X,W)\) after integrating out $W$.

Combining the two propensity score models and the two outcome regression models yields four scenarios, denoted \texttt{CC}, \texttt{CI}, \texttt{IC}, and \texttt{II}; the first and second letters indicate correct (\texttt{C}) or incorrect (\texttt{I}) specification of the propensity score and outcome regression working models, respectively.
Both specifications of the conditional mean in \eqref{eq:simulation-outcome-means} yield the optimal treatment rule $d_0(X)=\ind\{1-2X_1+X_2>0\}$.
In our simulations, we use the practical polynomial basis $h(X)=\{1,X_1,\allowbreak X_2,\allowbreak X_1^2,\allowbreak X_2^2,\allowbreak X_1X_2\}$. Each configuration uses 500 Monte Carlo replications. 
Supplementary Section~\ref{sec:app-auxiliary-efficiency} establishes that using $W_1$ and $W_2$ as additional predictors in the outcome component cannot increase asymptotic variance under certain conditions, while leaving the class of treatment rules unchanged. Supplementary Section~\ref{sec:app-simulation} reports additional simulations illustrating this efficiency comparison.

\begin{figure}[htbp]
\centering
\includegraphics[width=\textwidth]{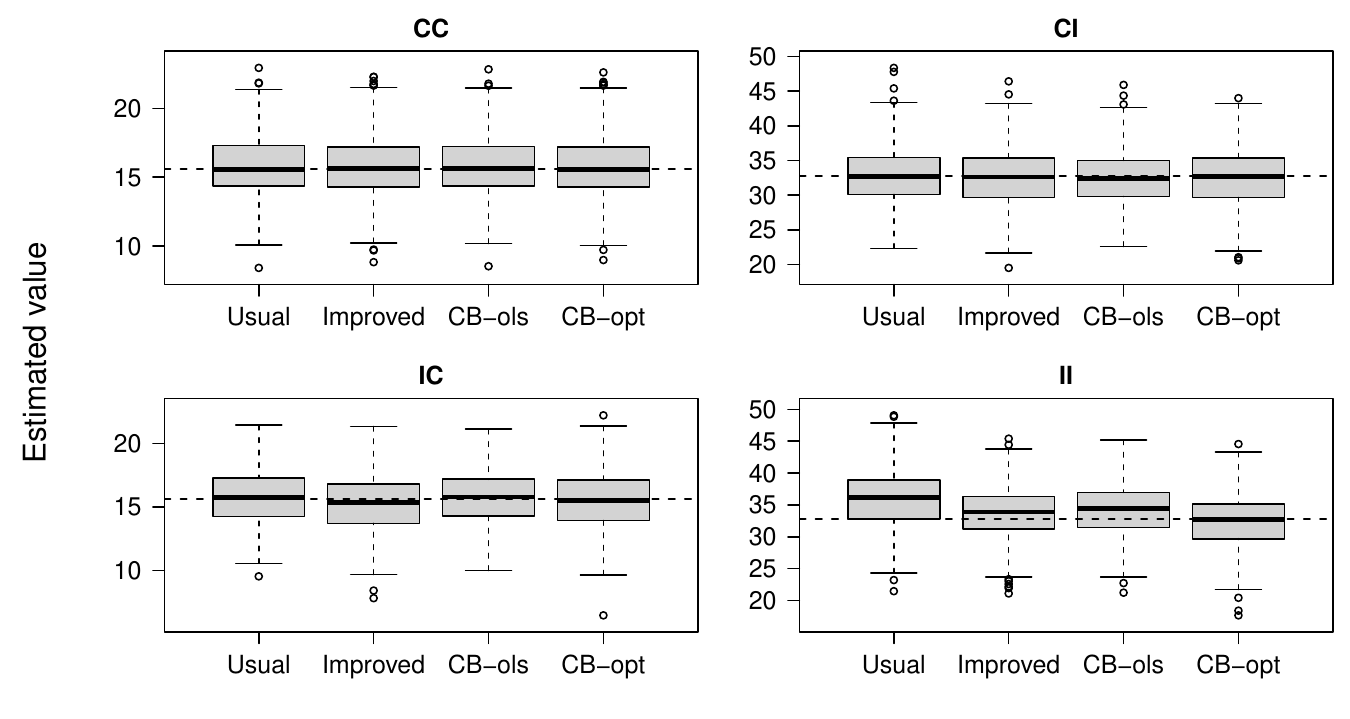}
\caption{Value estimates for the fixed rule $d_0$ at $n=1000$ across four scenarios. Each panel contains 500 Monte Carlo replications, and the dashed horizontal line is the true value.}
\label{fig:fixed-x-n1000}
\end{figure}

Figure~\ref{fig:fixed-x-n1000} compares the four estimators for the fixed rule $d_0$ at $n=1000$. In \texttt{CC}, \texttt{CI}, and \texttt{IC}, both \texttt{CB-opt} and \texttt{Usual} are essentially unbiased and have similar dispersion within each scenario, indicating that the proposed procedure preserves the favorable performance of the usual doubly robust estimator when at least one working model is correct. 
In \texttt{IC}, the proposed \texttt{CB-opt} has less bias than \texttt{Improved} (detailed results are reported in Table~\ref{tab:fixed-all-supp}), even though both methods use a variance-optimized outcome parameter. This pattern suggests that the chosen polynomial basis is sufficiently rich to control residual bias in this design. Supplementary Section~\ref{sec:app-simulation} reports sensitivity analyses using alternative bases.

Under joint misspecification (\texttt{II}), \texttt{CB-opt} has substantially less bias than the competing estimators while retaining comparable dispersion. The contrast between \texttt{CB-ols} and \texttt{CB-opt} isolates the role of outcome parameter selection because both use the same covariate balancing propensity score estimate: replacing the ordinary least squares outcome parameter with the parameter selected by the variance criterion markedly reduces the residual bias. Although this criterion directly targets variance, in this design it also yields a smaller bias interaction $\E\{\kappa(X)q_d(X)\}$, as shown in Supplementary Table~\ref{tab:fixed-II-population-diagnostics}. Thus, \texttt{CB-opt} combines competitive variability with protection against residual bias; Supplementary Section~\ref{sec:app-simulation} reports sensitivity analyses for the strength of misspecification and the choice of balancing basis.

We next estimate the optimal treatment rule by maximizing each value estimator over the linear rule class
\[
\mathcal D
=\left\{
d_{\eta}(X)=\ind\{\eta_0+\eta_1^\top X>0\}:
\eta=(\eta_0,\eta_1^\top)^\top,\ \|\eta\|_2=1
\right\}.
\]
For each scenario and $n\in\{500,1000,2000,3000,4000,5000\}$, we use 500 replications and evaluate the regret of the learned rule $\widehat d$ by $\mathcal R(\widehat d)=V(d^{\opt})-V(\widehat d)$.
The computational parameterization, optimization, and detailed numerical results are described in Supplementary Section~\ref{sec:app-learned-implementation}.

\begin{figure}[htbp]
\centering
\includegraphics[width=\textwidth]{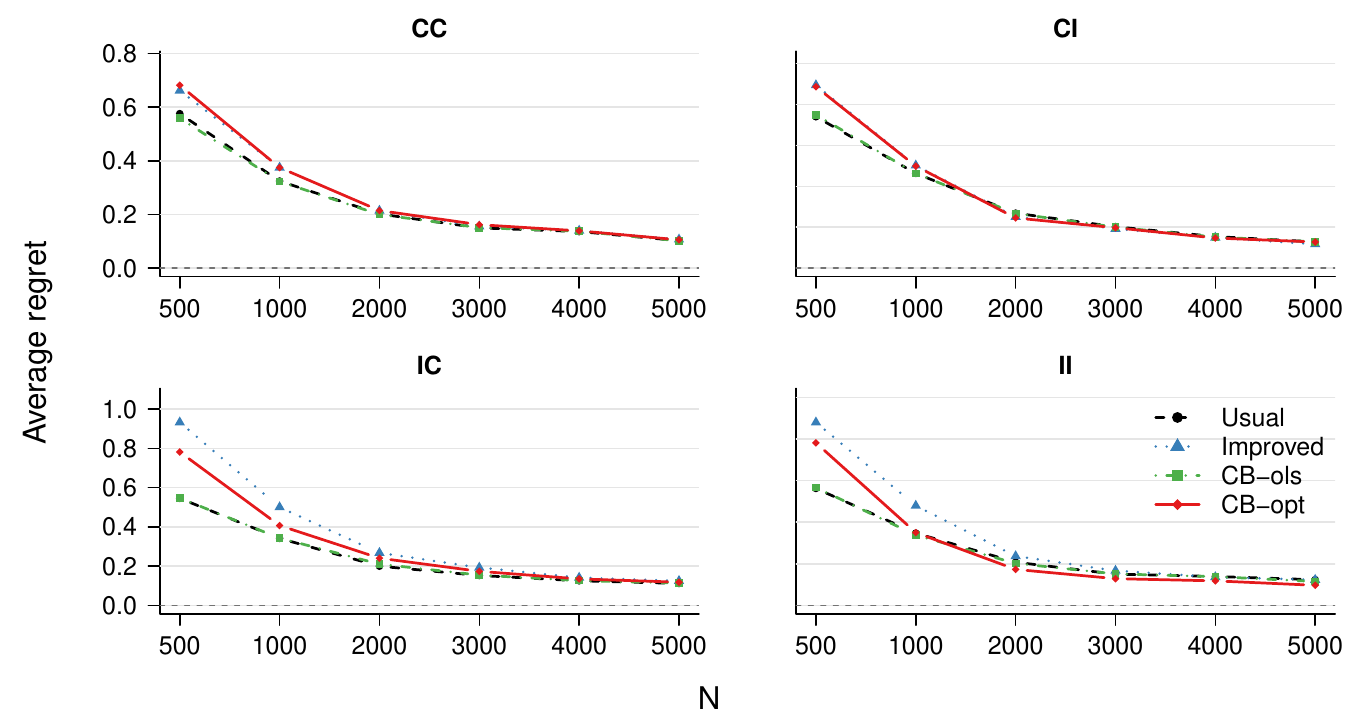}
\caption{Average regret of the learned rules across four scenarios. Each point averages 500 Monte Carlo replications; lower values indicate better performance.}
\label{fig:learned-X-all-main}
\end{figure}

Figure~\ref{fig:learned-X-all-main} shows that average regret decreases with sample size in all four scenarios, with the clearest advantage for \texttt{CB-opt} under joint misspecification (\texttt{II}). At $n=500$, the additional variability from estimating the outcome parameter by the variance criterion appears to outweigh the reduction in bias. From $n=2000$ onward, however, \texttt{CB-opt} has the smallest average regret, and its advantage persists as $n$ increases. This pattern agrees with previous results: once sampling variability diminishes, the smaller systematic error in value estimation leads to fewer errors in ranking candidate rules.

In \texttt{CC} and \texttt{IC}, a correctly specified outcome model makes the simpler \texttt{Usual} and \texttt{CB-ols} estimators sufficiently accurate for ranking rules, so variance optimization offers little additional benefit and can impose a cost in finite samples. In \texttt{CI}, the variance-optimized estimators \texttt{Improved} and \texttt{CB-opt} become competitive at moderate sample sizes, but their regret advantage is small and not uniform. These results are consistent with Proposition~\ref{prop:asymptotic-properties}(i), because pointwise variance optimality for each prespecified rule does not guarantee the most stable ranking over an entire rule class. Overall, the usual estimator can be adequate when a simple outcome model is credible or the sample is small, whereas \texttt{CB-opt} offers a useful safeguard when joint misspecification is plausible and the sample size is at least moderate.

\section{Application to the Leukemia Data}
\label{sec:application}

In this section, we apply our proposed strategy to an observational study of 1,161 patients with leukemia treated at Peking University People's Hospital between 2009 and 2017.
The clinical objective is to compare two types of hematopoietic stem-cell transplantation with respect to the time to leukemia relapse.
We set $A=1$ for HLA-matched sibling-donor transplantation and $A=0$ for haploidentical stem-cell transplantation.
The outcome $Y$ is the number of days to relapse, so a larger value is preferable.

We select six baseline covariates: patient age ($X_1$), duration from diagnosis to transplantation ($X_2$), minimal residual disease level before transplantation ($X_3$), donor sex ($X_4$), patient sex ($X_5$), and acute lymphoblastic leukemia subtype ($X_6$). The first three covariates are continuous, and the remaining three are binary.
The continuous covariates are normalized before analysis.
These variables describe related characteristics available before transplantation and provide a clinically interpretable adjustment set for comparing treatment strategies and constructing individualized decisions.
Because transplantation choices in this observational study may depend on baseline patient, disease, and donor characteristics, evaluating individualized treatment rules requires adjustment for these systematic differences. This motivates the use of our proposed covariate balancing doubly robust estimator.

We randomly split the data into a training sample of 580 patients and a test sample of 581 patients.
For each method, the treatment rule is learned using the training sample and its value is then estimated using the independent test sample.
We compare \texttt{CB-opt} with the three benchmarks considered in Section~\ref{sec:simulation}: \texttt{Usual}, \texttt{Improved}, and \texttt{CB-ols}.
The propensity score working model is a logistic regression on the six baseline covariates, and the outcome working model includes treatment, the six covariates, and all interactions between treatment and the covariates.
For \texttt{CB-ols} and \texttt{CB-opt}, we use the prespecified linear balancing basis $h(X)=\{1,X^\top\}^\top$.

\begin{table}[htbp]
\caption{Value estimates, standard errors, and 95\% confidence intervals for the leukemia data.}
\label{tab:leukemia-application}
\centering
\begin{tabular}{@{}cccccc@{}}
\toprule
Method & Point estimate & IF SE$^1$ & IF 95\% CI$^1$ & Bootstrap SE$^2$ & Bootstrap 95\% CI$^2$ \\
\midrule
\texttt{Usual}    & 1588 & 92 & $(1408, 1768)$ & 103 & $(1386, 1790)$ \\
\texttt{Improved} & 1606 & 86 & $(1438, 1775)$ & 112 & $(1387, 1826)$ \\
\texttt{CB-ols}   & 1594 & 96 & $(1406, 1782)$ & 105 & $(1388, 1800)$ \\
\texttt{CB-opt}   & 1626 & 90 & $(1450, 1803)$ & 115 & $(1401, 1851)$ \\
\bottomrule
\end{tabular}%
\par\smallskip
\begin{minipage}{0.975\textwidth}
\small
\noindent $^1$ The influence-function standard error and 95\% CI are conditional on the rule learned from the training sample.
\par\noindent $^2$ The bootstrap standard error and normal-approximation 95\% CI use 300 bootstrap resamples of the test sample, also conditional on the learned rule; the nuisance models are refitted in each resample.
\end{minipage}
\end{table}

Table~\ref{tab:leukemia-application} reports the estimated values and corresponding uncertainty measures. The estimated value of the \texttt{CB-opt} rule is 1626 days, or approximately 4.5 years, and its point estimate is the largest among the four methods. 
The confidence intervals across the four methods overlap substantially. The influence-function and bootstrap standard errors for \texttt{CB-opt} are of the same general order as those of the competing methods.
Moreover, the \texttt{CB-opt} rule recommends HLA-matched sibling-donor transplantation to 46.8\% of patients in the test sample and haploidentical transplantation to the remaining 53.2\%.
With the normalized coefficient vector, the fitted rule is \(\widehat d_{\texttt{CB-opt}}(X)=\ind\{0.449-0.089X_1-0.567X_2+0.245X_3-0.051X_4-0.604X_5-0.203X_6>0\}\).
Thus \texttt{CB-opt} yields an operational rule that is specific to each patient rather than a uniform treatment recommendation, while the independent test sample quantifies the expected mean time to relapse under that rule.
Supplementary Section~\ref{sec:app-leukemia-split-sensitivity} reports a descriptive sensitivity analysis over repeated random splits.

Covariate balance is improved under both propensity score estimators. In the test sample, the maximum absolute standardized mean difference across the six baseline covariates was 0.639 before weighting, 0.077 after weighting by the maximum-likelihood propensity score, and below 0.001 after weighting based on covariate balancing. Compared with maximum-likelihood weighting, the covariate-balancing weights nearly eliminate measured baseline imbalance, highlighting a key empirical advantage of the proposed approach for value estimation in this dataset.

\section{Discussion}
\label{sec:discussion}

This paper develops a covariate balancing doubly robust framework for value estimation and learning optimal individualized treatment rules, combining balance of covariate functions relevant to the rule with selection of the outcome regression parameter using a variance criterion.
A distinctive feature is its robustness characterization under joint model misspecification: consistency can still hold when the rule-relevant outcome regression error lies in the effective balancing space, even if neither working model is correctly specified.
The framework also provides variance optimality within the proposed class, semiparametric efficiency when both working models are correctly specified, regret guarantees for learned rules, and an extension using auxiliary predictors that can improve the efficiency of value estimation without enlarging the set of variables used by the treatment rule.

Because the rule-relevant outcome regression error depends on unknown population quantities, the span condition is generally not directly verifiable or guaranteed by the polynomial bases used in practice; performance therefore depends on how well the chosen basis approximates the relevant directions. Enriching the basis may reduce residual bias but can increase weight variability under limited overlap; approximate balance or regularized basis selection may therefore provide a useful trade-off between bias and variance \citep{WangZubizarreta2020}. A complementary direction is to use the outcome when constructing the basis. Cross-balancing separates feature construction from weight estimation through sample splitting \citep{JinZubizarreta2025} and may enlarge the effective balancing space while preserving valid inference. Finally, inference for the value of a learned optimal rule may be nonregular when the optimum is nonunique or treatment effects are close to zero \citep{LuedtkeVanderLaan2016}; extending inference under covariate balancing to this setting remains an important direction.

\section*{Declaration of generative AI and AI-assisted technologies in the manuscript preparation process}

During the preparation of this work, the authors used OpenAI ChatGPT to improve the language, readability, and organization of the manuscript and to check its consistency. After using these tools, the authors reviewed and edited the content as needed and take full responsibility for the content of the published article.

\clearpage
\bibliographystyle{apalike}
\bibliography{mybib_v2}

\clearpage
\section*{Supplementary Material}
\addcontentsline{toc}{section}{Supplementary Material}
\setcounter{equation}{0}
\renewcommand{\theequation}{S\arabic{equation}}
\renewcommand{\theHequation}{S\arabic{equation}}
\setcounter{lemma}{0}
\renewcommand{\thelemma}{S\arabic{lemma}}
\renewcommand{\theHlemma}{S\arabic{lemma}}
\setcounter{assumption}{0}
\renewcommand{\theassumption}{S\arabic{assumption}}
\renewcommand{\theHassumption}{S\arabic{assumption}}
\setcounter{theorem}{0}
\renewcommand{\thetheorem}{S\arabic{theorem}}
\renewcommand{\theHtheorem}{S\arabic{theorem}}
\setcounter{proposition}{0}
\renewcommand{\theproposition}{S\arabic{proposition}}
\renewcommand{\theHproposition}{S\arabic{proposition}}
\setcounter{remark}{0}
\renewcommand{\theremark}{S\arabic{remark}}
\renewcommand{\theHremark}{S\arabic{remark}}
\setcounter{table}{0}
\renewcommand{\thetable}{S\arabic{table}}
\renewcommand{\theHtable}{S\arabic{table}}
\setcounter{figure}{0}
\renewcommand{\thefigure}{S\arabic{figure}}
\renewcommand{\theHfigure}{S\arabic{figure}}

Supplementary Section~\ref{sec:app-auxiliary-efficiency} develops the framework with auxiliary predictors and establishes the resulting asymptotic variance comparison.
Supplementary Section~\ref{sec:app-regularity} provides additional regularity conditions for consistency and inference, together with sufficient conditions for recovering consistency under correct outcome regression specification.
Supplementary Section~\ref{sec:regret} gives a regret bound for estimated optimal rules.
Supplementary Section~\ref{sec:app-simulation} reports additional simulations and sensitivity analyses.
Supplementary Section~\ref{sec:app-proofs} provides the proofs.

\appendix
\section{Auxiliary Predictors and Asymptotic Variance Comparison}
\label{sec:app-auxiliary-efficiency}

We now incorporate auxiliary predictors $W$ into the framework developed in the main text. These covariates are predictive of the outcome but are not affected by treatment, as shown in Figure~\ref{fig:auxiliary_predictor}. In this section, write $O=(Y,A,X,W)$ with a slight abuse of notation.
Define $Z=(X^\top,W^\top)^\top.$
Throughout this analysis, the final decision rule depends only on $X$; the auxiliary predictors $W$ are not used as decision variables.
This restriction ensures that the rule can be implemented using the baseline decision covariates available at treatment assignment.
In addition to Assumption~\ref{assump:regular}, the analysis with auxiliary predictors imposes the following restriction.

\begin{assumption}\label{assump:auxiliary-regular}
$A \indep \{W,Y(1),Y(0)\} \mid X $.
\end{assumption}

Assumption~\ref{assump:auxiliary-regular} strengthens Assumption~\ref{assump:regular}(iii) by requiring treatment to be jointly independent of $W$ and the potential outcomes given $X$.
Thus, the predictors in $W$ may be associated with the outcome and with $X$, but they provide no additional information about treatment assignment once $X$ is given.
Moreover, $e_a^{\true}(Z)=e_a^{\true}(X)$ under Assumption~\ref{assump:auxiliary-regular}, where $e_a^{\true}(L)=\Prb(A=a\mid L)$ denotes the true propensity score for $L\in\{X,Z\}$.
Under Assumptions~\ref{assump:regular} and~\ref{assump:auxiliary-regular}, either $L=X$ or $L=Z$ identifies the same value $V(d)$, while $Z$ may provide additional predictive information for more efficient value estimation.

\begin{remark}
The framework also accommodates partially observed $W$, for example when these predictors are costly to measure or unavailable at treatment assignment.
Let $R$ be the vector of observation indicators for the components of $W$, and let $g$ be a deterministic vector function of $X$ of the same dimension as $W$.
One may replace $W$ with $W_g=R\odot W+(\mathbf 1-R)\odot g(X)$, where $\odot$ denotes elementwise multiplication, and then use $Z_g=(X^\top,W_g^\top,R^\top)^\top$ as the auxiliary vector.
The observation mechanism may depend on $W$, provided that $R\indep\{A,Y(1),Y(0)\}\mid(X,W)$.
For simplicity, we focus here on completely observed $W$.
\end{remark}

\begin{figure}[htbp]
\centering
    \begin{tikzpicture}[>=stealth, scale=1.2]
        \node[draw, circle, inner sep=0pt, minimum size=0.8cm, line width=0.7pt] (XP) at (1, 1.8) {$W$};
        \node[draw, circle, inner sep=0pt, minimum size=0.8cm, line width=0.7pt] (XC) at (-1, 1.8) {$X$};
        \node[draw, circle, inner sep=0pt, minimum size=0.8cm, line width=0.7pt] (A) at (-1, 0) {$A$};
        \node[draw, circle, inner sep=0pt, minimum size=0.8cm, line width=0.7pt] (Y) at (1, 0) {$Y$};
        
        \draw[->, line width=0.7pt] (XC) -- (A);
        \draw[->, line width=0.7pt] (XC) -- (Y);
        \draw[->, line width=0.7pt] (XP) -- (Y);
        \draw[->, line width=0.7pt] (A) -- (Y);
        \draw[->, line width=0.7pt] (XC) -- (XP);
    \end{tikzpicture}
\caption{A causal diagram illustrating the auxiliary predictors $W$ under Assumption~\ref{assump:auxiliary-regular}.}
\label{fig:auxiliary_predictor}
\end{figure}

For the efficiency comparison, we refer to the procedures whose outcome augmentation uses \(L=X\) and \(L=Z\) as the \(X\)-based and \(Z\)-based analyses, respectively. A subscript $L\in\{X,Z\}$ denotes the corresponding quantity from the main text obtained using covariate set $L$, while the treatment rule remains $d(X)$.
Under correct propensity score specification, write the corresponding asymptotic variance as
\begin{equation*}
\mathcal S_{\subL}(\beta_{\subL}) =\mathcal S(\alpha^{\true},\beta_{\subL}) =\E\left[ \{\varphi_{\subL}^{\cb}(O;\alpha^{\true},\beta_{\subL})\}^2 \right],
\end{equation*}
and define $\beta_{\subL}^{\opt}\in\argmin_{\beta_{\subL}}\mathcal S_{\subL}(\beta_{\subL}).$
Let $\widehat\beta_{\subL}^{\opt}$ denote the corresponding empirical minimizer.

\begin{assumption}
\label{assump:nested-first-order}
Suppose the following conditions hold for the $X$-based and $Z$-based analyses: 
\begin{enumerate}[label=(\roman*)]
\item Both analyses use a common, correctly specified propensity score model based on $X$, that is, $e_a(X;\alpha^{\true})=e_a^{\true}(X)$ for $a\in\{0,1\}$.
\item The augmentation model based on $Z$ nests the model based on $X$, that is, for every $\beta_{\subX}$, there exists $\beta_{\subZ}$ such that $m_a(Z;\beta_{\subZ})=m_a(X;\beta_{\subX})$ almost surely for $a\in\{0,1\}$.
\item Both analyses use the same balancing basis $h(X;\alpha)$.
\end{enumerate}
\end{assumption}

Under Assumption~\ref{assump:auxiliary-regular}, Assumption~\ref{assump:nested-first-order}(i) can be satisfied without including the predictors in $W$ in the propensity score model. Assumption~\ref{assump:nested-first-order}(ii) holds, for example, when the model based on $Z$ adds main effects and treatment interactions for the predictors in $W$ to the model based on $X$, since setting the corresponding coefficient vectors to zero recovers the latter. 
For example, if $m_a(X;\beta_{\subX})=\beta_{\subX,0}+\beta_{\subX,1}a+\beta_{\subX,2}^\top X+a\beta_{\subX,3}^\top X$, a linear model based on $Z$ can be chosen as $m_a(Z;\beta_{\subZ})=\beta_{\subZ,0}+\beta_{\subZ,1}a+\beta_{\subZ,2}^\top X+a\beta_{\subZ,3}^\top X+\beta_{\subZ,4}^\top W+a\beta_{\subZ,5}^\top W$; setting $\beta_{\subZ,4}=\beta_{\subZ,5}=0$ recovers the $X$-based model.
Assumption~\ref{assump:nested-first-order}(i) and (iii) together hold the propensity score estimator fixed across the two analyses.
For the efficiency result below, we additionally assume that \(\beta_{\subL}^{\opt}\) exists and that \(\widehat\beta_{\subL}^{\opt}\overset{p}{\to}\beta_{\subL}^{\opt}\) for $L\in\{X,Z\}$. 

\begin{proposition}
\label{prop:z-efficiency}
Suppose Assumptions~\ref{assump:auxiliary-regular} and \ref{assump:nested-first-order} hold.
Then the asymptotic variance of the covariate balancing doubly robust estimator based on $Z$ is no larger than that of the estimator based on $X$:
\begin{equation}\label{eq:variance-ordering}
\mathcal S_{\subZ}(\beta_{\subZ}^{\opt}) \leq \mathcal S_{\subX}(\beta_{\subX}^{\opt}).
\end{equation}
\end{proposition}

Proposition~\ref{prop:z-efficiency} shows that incorporating auxiliary predictors into the augmentation model cannot increase asymptotic variance while the treatment rule remains based only on \(X\). By Assumption~\ref{assump:nested-first-order}(ii), the augmentation class based on $Z$ contains every model based on $X$ and may also capture outcome variation explained by $W$. Because propensity score estimation is held fixed under Assumption~\ref{assump:nested-first-order}(i) and (iii), minimizing the variance criterion over this richer class yields an asymptotic variance no greater than that obtained from the class based on $X$. The gain can be strict when the additional predictors supply useful prognostic information beyond $X$.

This result highlights the value of prognostic covariates for causal estimation \citep{hahn2004functional,brookhart2006variable} and complements recent doubly robust efficiency results for nonconfounding predictive covariates \citep{luo2025efficiency}. Importantly, the improvement does not require correct specification of the outcome regression working models or expansion of the rule class. It is therefore particularly relevant when treatment assignment is known by design or can be credibly modeled using $X$, while additional predictors are available for outcome modeling. Supplementary Section~\ref{sec:app-aux-simulation} illustrates the efficiency gains for a fixed rule, and Supplementary Section~\ref{sec:app-learned-implementation} examines their implications for learned rules.

\section{Regularity Conditions and Consistency under a correct outcome model}
\label{sec:app-regularity}

\subsection{Regularity conditions for consistency and inference}
\label{sec:app-regularity-conditions}
\begin{assumption}\label{assump:app-cb-regularity}
Let $\Psi(\alpha)=\E\{\psi(A,X;\alpha)\}$ and $Q(\alpha)=\|\Psi(\alpha)\|_2^2$.
Define the sample analogues by $\widehat\Psi_n(\alpha)=\Pn\{\psi(A,X;\alpha)\},$ and $widehat Q_n(\alpha)=\|\widehat\Psi_n(\alpha)\|_2^2.$
Let \(\widehat\alpha^{\cb}\) minimize the sample analogue of $Q(\alpha)$ as defined in \eqref{eq:cb-estimating-equation}.
Suppose the following conditions hold:

\begin{enumerate}[label=(\roman*)]
\item The parameter space $\mathcal A$ of $\alpha$ is compact and contains the target point \(\alpha^*\) as an interior point.

\item The population criterion \(Q(\alpha)\) has a unique minimizer \(\alpha^*\).

\item The map \(\alpha\mapsto \Psi(\alpha)\) is continuously differentiable in a neighborhood of \(\alpha^*\), with $G^*=G(\alpha^*)=\E\{\partial\psi(A,X;\alpha)/\partial\alpha^\top\}|_{\alpha=\alpha^*}$.
The matrix \((G^*)^\top G^*\) is nonsingular.

\item For some neighborhood \(\mathcal N\) of \(\alpha^*\),
\[
\sup_{\alpha\in\mathcal A} \left\| (\Pn-\Pp)\psi(A,X;\alpha) \right\| =o_p(1), ~~ \sup_{\alpha\in\mathcal N} \left\| (\Pn-\Pp)\frac{\partial \psi(A,X;\alpha)} {\partial\alpha^\top} \right\| =o_p(1),
\]
where \(\Pp f=\E\{f(O)\}\) for any integrable function \(f\).
Moreover, $\E\|\psi(A,X;\alpha^*)\|^2<\infty$.

\end{enumerate}
\end{assumption}

Assumption~\ref{assump:app-cb-regularity} gives standard GMM conditions for consistency and asymptotic linearity of \(\widehat\alpha^{\cb}\) \citep{Hansen1982,newey1994large}. Under correct specification, they yield \eqref{eq:alpha-if}. When the propensity score model is misspecified, these conditions ensure that \(\widehat\alpha^{\cb}\) converges to the pseudo-true minimizer \(\alpha^*\) of \(Q(\alpha)\). Because \(\alpha^*\) is an interior minimizer, it satisfies the population first-order condition \eqref{eq:cb-population-foc}, which yields the orthogonality used in Proposition~\ref{prop:double-robustness}(ii). 

\emph{Additional maintained conditions.}
For each fixed rule, value consistency also requires $\widehat\beta\overset{p}{\to}\beta^*$ and the local Glivenko--Cantelli and continuity conditions for the value estimating function underlying \eqref{eq:cb-value-class}. Variance optimization and inference additionally require a unique interior minimizer that is well separated from other parameter values, together with the standard smoothness, moment, nonsingularity, and stochastic equicontinuity conditions needed for the joint $M$-estimation expansion and the influence function representation \eqref{eq:cb-value-if} \citep{newey1994large}.

\subsection{Recovering consistency under correct outcome regression specification}
\label{sec:app-or-route}

Suppose the outcome regression model is correctly specified at some interior parameter $\beta^{\true}$. Correct specification alone does not imply $\beta^*=\beta^{\true}$ because $\widehat\beta^{\opt}$ minimizes the variance criterion rather than a loss used to fit the outcome regression. Under strict convexity, the usual consistency route based on a correct outcome regression model is recovered if
\begin{equation}\label{eq:or-variance-minimizer}
\nabla_\beta\mathcal S(\alpha^*,\beta^{\true})=0.
\end{equation}
We now give a sufficient balancing condition for \eqref{eq:or-variance-minimizer}. The argument follows by differentiating the same second-moment criterion used to select $\widehat\beta^{\opt}$.

For $a\in\{0,1\}$, define the derivative vectors $e_{a,\alpha}(X;\alpha)=\partial e_a(X;\alpha)/\partial\alpha$ and $m_{a,\beta}(X;\beta)=\partial m_a(X;\beta)/\partial\beta$. Let $\widetilde\alpha$ and $\widetilde\beta$ be pilot estimators of the propensity score and outcome regression parameters, respectively, and define
\begin{equation}\label{eq:or-arm-balancing-functions}
H_a(X;\widetilde\alpha,\widetilde\beta)
=
\begin{pmatrix}
\{1-e_a(X;\widetilde\alpha)\}
 m_{a,\beta}(X;\widetilde\beta)\\
\{1-e_a(X;\widetilde\alpha)\}
 m_a(X;\widetilde\beta)
 m_{a,\beta}(X;\widetilde\beta)\\
e_{a,\alpha}(X;\widetilde\alpha)\\
m_a(X;\widetilde\beta)
 e_{a,\alpha}(X;\widetilde\alpha)
\end{pmatrix},
~~ a\in\{0,1\}.
\end{equation}
where $e_{d,\alpha}(X;\alpha)=d(X)e_{1,\alpha}(X;\alpha)+\{1-d(X)\}e_{0,\alpha}(X;\alpha),$ and $m_{d,\beta}(X;\beta)=d(X)m_{1,\beta}(X;\beta)+\{1-d(X)\}m_{0,\beta}(X;\beta).$
Consider the stacked balancing basis
\begin{equation}\label{eq:or-sufficient-balancing-functions}
h(X;\widetilde\alpha,\widetilde\beta)
=
\begin{pmatrix}
H_0(X;\widetilde\alpha,\widetilde\beta)\\
H_1(X;\widetilde\alpha,\widetilde\beta)
\end{pmatrix}.
\end{equation}
In implementation, $\widetilde\alpha$ and $\widetilde\beta$ are held fixed when \eqref{eq:or-sufficient-balancing-functions} is used in the criterion \eqref{eq:cb-estimating-equation}. Stacking the two treatment arms gives a common basis for different rules.

Let $h^*(X)$ be the probability limit of the basis in \eqref{eq:or-sufficient-balancing-functions}. In this section, use $h^*(X)$ in the definitions of $\psi$, $G^*$, and $\mathcal H_{\rm eff}$. Set $K^*=\{(G^*)^\top G^*\}^{-1}(G^*)^\top$ and $D^*=\partial\{\Gamma(\alpha^*,\beta)K^*\}/\partial\beta|_{\beta=\beta^{\true}}$. 
When the outcome model is correctly specified, a direct calculation gives
\begin{equation}\label{eq:or-variance-gradient-representation}
\nabla_\beta\mathcal S(\alpha^*,\beta^{\true})
=2\E\left[
\kappa(X)\{m_d(X;\beta^{\true})-V(d)\}b_d(X)
\right],
\end{equation}
where 
\begin{equation}\label{eq:or-gradient-direction}
b_d(X)=D^*h^*(X)
-\{2d(X)-1\}\{1-e_d(X;\alpha^*)\}
m_{d,\beta}(X;\beta^{\true}).
\end{equation}
Thus, it is enough for every component of $b_d$ and $m_d(\cdot;\beta^{\true})b_d$ to lie in the effective balancing space. This condition asks the GMM first-order condition to enforce only the directions that enter the variance gradient; it does not require exact population balance of the entire stacked basis.

Throughout this subsection, suppose that the pilot estimators have fixed probability limits, with $\widetilde\beta\overset{p}{\to}\beta^{\true}$. Also suppose that $\Gamma(\alpha^*,\beta)$ is differentiable at $\beta^{\true}$ and that $\mathcal S(\alpha^*,\beta)$ is differentiable and strictly convex in $\beta$ on a convex parameter space. The regularity conditions in Supplementary Section~\ref{sec:app-regularity-conditions} are maintained.

\begin{proposition}
\label{prop:or-route-sufficient}
Fix a treatment rule $d$ and use the basis in \eqref{eq:or-sufficient-balancing-functions}. Suppose the outcome regression model is correctly specified, so that $m_a(X;\beta^{\true})=m_a^{\true}(X)$ for $a\in\{0,1\}$. If every component of $b_d(X)$ and $m_d(X;\beta^{\true})b_d(X)$ belongs to $\mathcal H_{\rm eff}$, then $\beta^*=\beta^{\true}$ and $\widehat V^{\cb}(d)\overset{p}{\to}V(d)$.
\end{proposition}

When the outcome regression model is correct, $\Gamma(\alpha^*,\beta^{\true})=0$. Its derivative with respect to $\beta$ may still be nonzero. In \eqref{eq:or-gradient-direction}, the term that contains $m_{d,\beta}$ comes from differentiating the doubly robust influence function. The term $D^*h^*$ comes from estimating the propensity score parameter by GMM. If the effective balancing space contains these directions, then $\beta^{\true}$ satisfies the first-order condition for minimizing the variance. Strict convexity then gives $\beta^*=\beta^{\true}$, and the value estimator is consistent.

Estimation with likelihood scores gives a simple special case. In that case, the last term in the influence function is a constant linear transformation of the likelihood score for the propensity score parameter. This gives the four groups of functions in \eqref{eq:or-arm-balancing-functions}. For a GMM estimator with more moment conditions than parameters, the required directions are those in \eqref{eq:or-gradient-direction}. Equation~\eqref{eq:or-sufficient-balancing-functions} therefore gives a possible choice of basis functions. Proposition~\ref{prop:or-route-sufficient} also requires these directions to belong to the effective balancing space. This requirement does not follow from the choice of basis functions alone. Cross-fitting can use separate samples to estimate the pilot quantities and evaluate the balancing functions. Supplementary Figure~\ref{fig:fixed-derivative-ic-supp} compares this theory-guided construction with a simpler polynomial basis when the outcome regression model is correct.

\section{Regret Bounds for the Learned Optimal Rule}
\label{sec:regret}

This section establishes a regret bound for the rule learned by maximizing the proposed value estimator based on observational data \citep{athey2021policy}.
The complexity of the rule class is allowed to grow with $n$ and we write the candidate rule class as $\mathcal D_n$.
Let $v_n=\operatorname{VC}(\mathcal D_n)$ denote its Vapnik--Chervonenkis (VC) dimension, $d_n^{\opt}\in\argmax_{d\in\mathcal D_n}V(d)$, and $\widehat d_n\in\mathcal D_n$ denote the learned rule.
The outcome parameter can vary with the rule, so we use the notation $\widehat\beta_d^{\opt}$. For each candidate rule, write $\widehat V^{\cb}(d)=\Pn\ell_d(O;\widehat\alpha^{\cb},\widehat\beta_d^{\opt})$, where
\begin{equation}\label{eq:rule-value-function}
\ell_d(O;\alpha,\beta)
=m_d(X;\beta)
+\frac{\ind\{A=d(X)\}}{e_d(X;\alpha)}
\left\{Y-m_d(X;\beta)\right\}.
\end{equation}
The quantity $\ell_d(O;\alpha,\beta)$ is referred to as the value estimating function for rule $d$ because its empirical average gives the estimated value of that rule after $\alpha$ and $\beta$ are replaced by their estimates.
Let $\alpha^*$ denote the probability limit of $\widehat\alpha^{\cb}$. 
For each rule $d$, let $\beta_d^*$ denote the probability limit of $\widehat\beta_d^{\opt}$. 
These limits need not equal the true parameters when the corresponding working model is misspecified. Define $\ell_d^*(O)=\ell_d(O;\alpha^*,\beta_d^*)$ as the value estimating function evaluated at these probability limits.
The uniform population value discrepancy is
\begin{equation}\label{eq:uniform-value-bias}
b_n =\sup_{d\in\mathcal D_n} \left|\E\left\{\ell_d^*(O)\right\}-V(d)\right|.
\end{equation}
Thus, $b_n$ is the largest residual population bias in value estimation over $\mathcal D_n$.
Here $q_d$ is defined as in Section~\ref{sec:balancing-space-bias}, with $\beta^*$ replaced by the limit $\beta_d^*$ for that rule. Equation~\eqref{eq:dr-value-decomp} gives two cases in which $b_n=0$. If the propensity score model is correctly specified, then $\kappa(X)=0$, so the population bias is zero for every rule in $\mathcal D_n$. If the propensity score model is misspecified, the population bias is still zero when $q_d\in\mathcal H_{\rm eff}$ for every $d\in\mathcal D_n$. More generally, the approximation bound in Section~\ref{sec:balancing-space-bias} gives
\[
b_n
\leq
\|\kappa\|_{L^2(P)}
\sup_{d\in\mathcal D_n}
\inf_{f\in\mathcal H_{\rm eff}}
\|q_d-f\|_{L^2(P)}.
\]
Thus, $b_n$ is small when every $q_d$ can be approximated well by the effective balancing space.

The following assumption controls four sources of error uniformly over the candidate rules: sampling variation, nuisance parameter estimation, residual population bias, and imperfect maximization of the empirical value. The first two conditions can be established using empirical process results for policy classes and uniform results for nuisance estimation or cross-fitting \citep{athey2021policy,Chernozhukov2018DML}.

\begin{assumption}
\label{assump:uniform-rule-learning}
Suppose the following conditions hold:
\begin{enumerate}[label=(\roman*)]
\item The class of value estimating functions evaluated at the probability limits satisfies
\[
\sup_{d\in\mathcal D_n}
\left|(\Pn-\Pp)\ell_d^*(O)\right|
=O_p\left(\sqrt{\frac{v_n}{n}}\right).
\]
\item For deterministic sequences $r_{\alpha,n},r_{\beta,n}\rightarrow0$, the error from estimating $\alpha$ and $\beta_d$ satisfies
\[
\sup_{d\in\mathcal D_n}
\left|\Pn\left\{\ell_d(O;\widehat\alpha^{\cb},\widehat\beta_d^{\opt})
-\ell_d^*(O)\right\}\right|
=O_p(r_{\alpha,n}+r_{\beta,n}).
\]
\item The uniform residual value bias in \eqref{eq:uniform-value-bias} satisfies $b_n\to0$.
\item For a deterministic sequence $\rho_n\geq0$ with $\rho_n\to0$, $\widehat d_n$ satisfies
\[
\widehat V^{\cb}(\widehat d_n) \geq \sup_{d\in\mathcal D_n}\widehat V^{\cb}(d)-\rho_n.
\]
\end{enumerate}
\end{assumption}

\begin{theorem}
\label{thm:regret-rate}
Suppose Assumptions~\ref{assump:regular} and \ref{assump:uniform-rule-learning} hold.
Then
\begin{equation*}
0 \leq V(d_n^{\opt})-V(\widehat d_n) =O_p\left( \sqrt{\frac{v_n}{n}} +r_{\alpha,n}+r_{\beta,n}+b_n+\rho_n \right).
\end{equation*}
\end{theorem}

Theorem~\ref{thm:regret-rate} shows how treatment rule complexity, nuisance estimation, residual value bias, and numerical optimization determine the regret rate. The bound is $O_p(n^{-1/2})$ if $v_n=O(1)$ and $r_{\alpha,n}$, $r_{\beta,n}$, $b_n$, and $\rho_n$ are all $O(n^{-1/2})$. The conditions $b_n=0$ and $\rho_n=0$ are sufficient special cases. For example, suppose $\mathcal X\subseteq\mathbb R^p$ and $p$ is fixed. The linear rule class used in Section~\ref{sec:simulation} has VC dimension $p+1$, and a decision tree class has fixed VC dimension when its depth is fixed. If the tree depth instead grows as \(L_n=\lfloor\gamma\log_2(n)\rfloor\) for some \(\gamma<1/2\), then $v_n=O(n^{1/2})$, and the regret bound is $o_p(1)$ \citep{athey2021policy}.

\section{Additional Numerical Results}
\label{sec:app-simulation}

\subsection{Efficiency gains from auxiliary predictors}\label{sec:app-aux-simulation}

\begin{figure}[htbp]
\centering
\includegraphics[width=\textwidth]{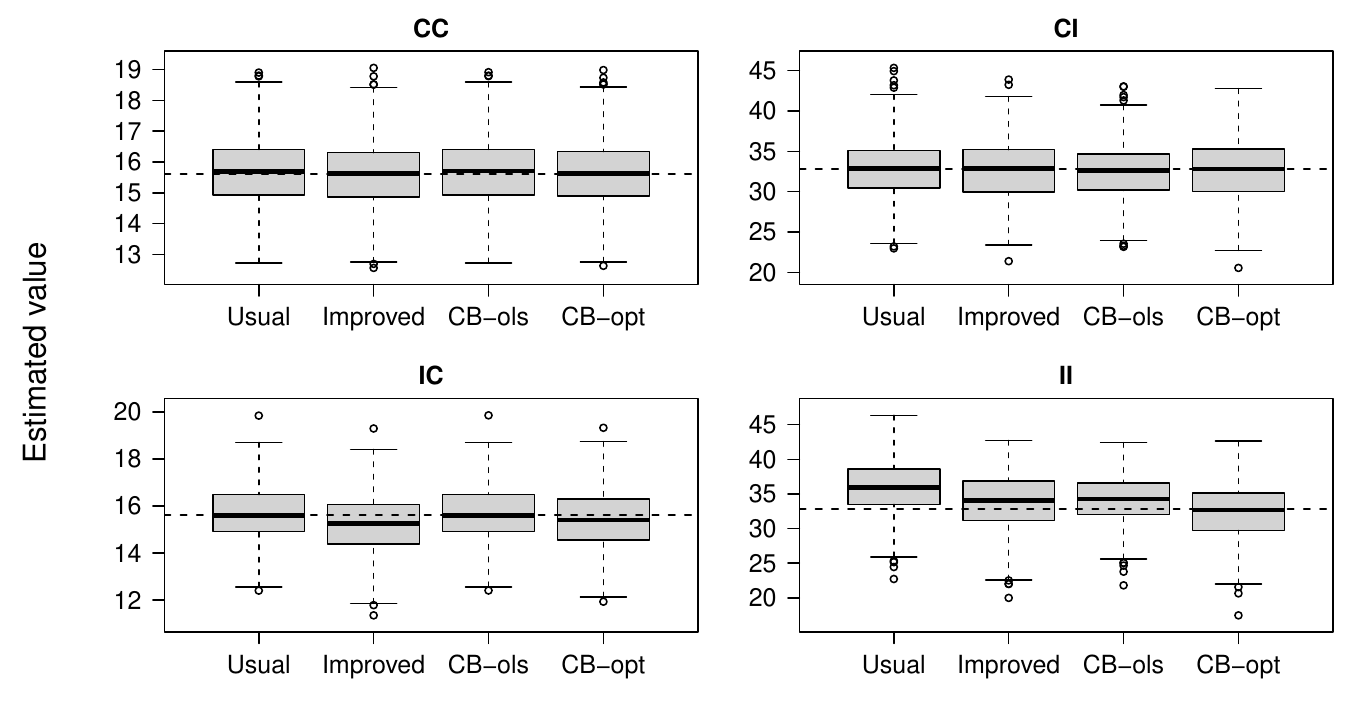}
\caption{Value estimates for the fixed rule $d_0$ based on $Z$ at $n=1000$. Dashed lines represent the true value.}
\label{fig:fixed-z-n1000-supp}
\end{figure}

Figure~\ref{fig:fixed-z-n1000-supp} reports the corresponding results after adding $W$ to the outcome augmentation, so that the analysis is based on $Z=(X^\top,W^\top)^\top$ while the rule remains $d_0(X)$. 
Table~\ref{tab:fixed-all-supp} complements Figure~\ref{fig:fixed-x-n1000} in the main text and includes the corresponding numerical results for the $Z$-based analysis. Its last column reports the variance reduction ratio $1-\widehat{\operatorname{Var}}\{\widehat V_Z(d_0)\}/\widehat{\operatorname{Var}}\{\widehat V_X(d_0)\},$ where each empirical variance is computed across 500 Monte Carlo replications. A positive value indicates that adding $W$ reduces the empirical variance.

For both the $X$- and $Z$-based analyses, \texttt{Usual}, \texttt{CB-ols}, and \texttt{CB-opt} have small bias when at least one working model is correct, whereas \texttt{Improved} has somewhat larger bias when only the outcome regression model is correct. Both \texttt{CB-opt} and \texttt{Improved} select the outcome parameter by minimizing the empirical variance, but \texttt{CB-opt} has smaller absolute bias than \texttt{Improved} across all four scenarios and for both covariate sets, suggesting that covariate balancing contributes to the bias reduction. In particular, \texttt{CB-opt} has the smallest absolute bias when both nuisance models are misspecified. 

When the propensity score model is correctly specified, as in scenarios \texttt{CC} and \texttt{CI}, the variance reduction ratio for \texttt{CB-opt} is positive. This is consistent with Proposition~\ref{prop:z-efficiency}, which allows the outcome regression working model to be misspecified. The reductions are larger across all four methods in scenarios \texttt{CC} and \texttt{IC}, where the outcome regression model is correct, than in scenario \texttt{CI}, where only the propensity score model is correct. The gain in scenario \texttt{IC} is not guaranteed by Proposition~\ref{prop:z-efficiency}, because the propensity score model is misspecified in this scenario. These results show that prognostic information in $W$ can improve efficiency without enlarging the rule class.

Under joint misspecification, Proposition~\ref{prop:z-efficiency} does not apply. In scenario \texttt{II}, the variance reduction ratio is negative for \texttt{Improved} but positive for the other three methods, so adding $W$ does not uniformly improve efficiency across methods. Nevertheless, \texttt{CB-opt} continues to have the smallest absolute bias, whereas the other methods retain larger biases. Thus, the effect of auxiliary predictors is mainly on efficiency improvement, but it need not be beneficial for every estimator; the choice of estimator remains important for robustness to joint misspecification.

\begin{table}[H]
\centering
\caption{Bias and Monte Carlo standard deviation for the $X$- and $Z$-based analyses, and the variance reduction ratio from incorporating the auxiliary predictors, for the fixed rule $d_0$ at $n=1000$.}
\label{tab:fixed-all-supp}
\begin{tabular}{llccc}
\toprule
\raisebox{1ex}[0pt][0pt]{Scenario}
& \raisebox{1ex}[0pt][0pt]{Method}
& \raisebox{1ex}[0pt][0pt]{$X$-based bias (MCSD)}
& \raisebox{1ex}[0pt][0pt]{$Z$-based bias (MCSD)}
& \shortstack{Variance reduction\\ratio} \\
\midrule
\multirow{4}{*}{\texttt{CC}}
& \texttt{Usual}    &  0.156 (2.225) &  0.063 (1.128) &  0.743 \\
& \texttt{Improved} &  0.097 (2.270) & $-0.032$ (1.154) &  0.741 \\
& \texttt{CB-ols}   &  0.154 (2.198) &  0.063 (1.128) &  0.737 \\
& \texttt{CB-opt}   &  0.092 (2.278) & $-0.014$ (1.155) &  0.743 \\
\addlinespace
\multirow{4}{*}{\texttt{CI}}
& \texttt{Usual}    &  0.030 (4.265) &  0.083 (3.634) &  0.274 \\
& \texttt{Improved} & $-0.197$ (4.267) & $-0.114$ (3.731) &  0.235 \\
& \texttt{CB-ols}   & $-0.279$ (4.065) & $-0.229$ (3.430) &  0.288 \\
& \texttt{CB-opt}   & $-0.165$ (4.204) & $-0.104$ (3.660) &  0.242 \\
\addlinespace
\multirow{4}{*}{\texttt{IC}}
& \texttt{Usual}    &  0.114 (2.098) &  0.071 (1.128) &  0.711 \\
& \texttt{Improved} & $-0.327$ (2.226) & $-0.392$ (1.212) &  0.704 \\
& \texttt{CB-ols}   &  0.112 (2.027) &  0.071 (1.128) &  0.690 \\
& \texttt{CB-opt}   & $-0.149$ (2.239) & $-0.194$ (1.210) &  0.708 \\
\addlinespace
\multirow{4}{*}{\texttt{II}}
& \texttt{Usual}    &  3.204 (4.592) &  3.199 (3.869) &  0.290 \\
& \texttt{Improved} &  0.986 (4.122) &  1.067 (4.140) & $-0.009$ \\
& \texttt{CB-ols}   &  1.479 (4.033) &  1.473 (3.493) &  0.250 \\
& \texttt{CB-opt}   & $-0.386$ (4.260) & $-0.321$ (4.056) &  0.094 \\
\bottomrule
\end{tabular}
\end{table}

\subsection{Robustness to nuisance model misspecification}

For \(L\in\{X,Z\}\), let \(m_a^{\true}(L)=\E\{Y(a)\mid L\}\), and let \(\beta_{d,L}^*\) denote the probability limit of the \(L\)-based outcome regression parameter for rule \(d\). Define the corresponding rule-relevant outcome regression error by $q_{d,L}(L)=d(X)\{1-e_1(X;\alpha^*)\}\{m_1^{\true}(L)-m_1(L;\beta_{d,L}^*)\}-\{1-d(X)\}e_1(X;\alpha^*)\{m_0^{\true}(L)-m_0(L;\beta_{d,L}^*)\}.$
Table~\ref{tab:fixed-II-population-diagnostics} reports the population bias interaction \(\E\{\kappa(X)q_{d_0,L}(L)\}\) in scenario \texttt{II}.
The smaller absolute bias interaction for \texttt{CB-opt} supports the corresponding conclusion in Section~\ref{sec:simulation} and provides evidence for the mechanism through which covariate balancing reduces residual bias.

\begin{table}[htbp]
\centering
\caption{Population bias interaction \(\E\{\kappa(X)q_{d_0,L}(L)\}\) in scenario \texttt{II}.}
\label{tab:fixed-II-population-diagnostics}
\begin{tabular}{lcccc@{\hspace{18pt}}cccc}
\toprule
\multirow{2}{*}{Scenario}
& \multicolumn{4}{c}{$X$-based}
& \multicolumn{4}{c}{$Z$-based} \\
\cmidrule(lr){2-5}\cmidrule(lr){6-9}
& \texttt{Usual} & \texttt{Improved} & \texttt{CB-ols} & \texttt{CB-opt}
& \texttt{Usual} & \texttt{Improved} & \texttt{CB-ols} & \texttt{CB-opt} \\
\midrule
\texttt{II}
& 3.205 & 1.893 & 1.630 & 0.188
& 3.018 & 1.200 & 1.618 & $-0.184$ \\
\bottomrule
\end{tabular}
\end{table}

\begin{figure}[htbp]
\centering
\includegraphics[width=0.8\textwidth]{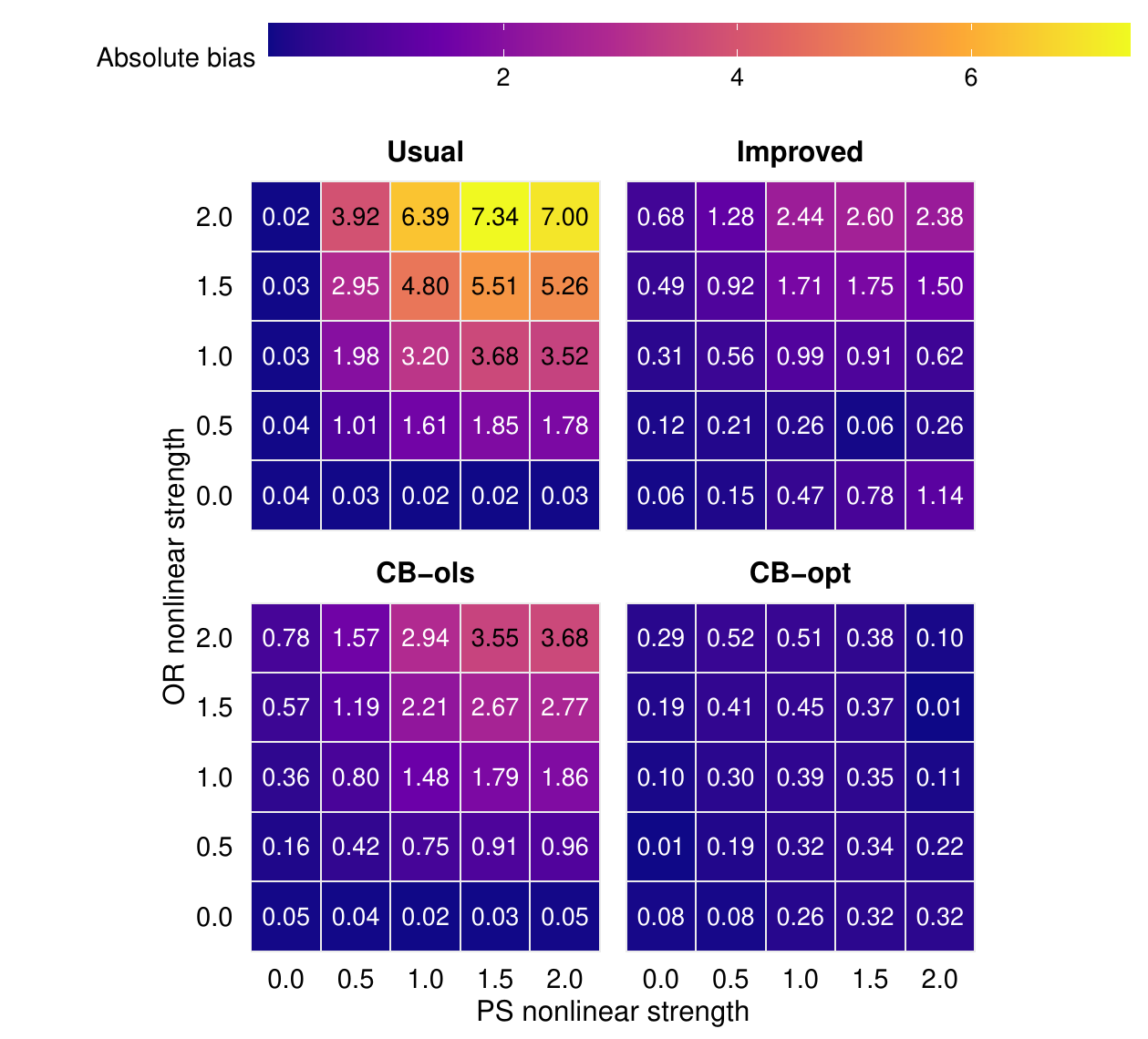}
\caption{Absolute bias for rule $d_0$ over a grid of nonlinear strengths in the propensity score (PS) and outcome regression (OR) at $n=1000$. Darker cells indicate smaller bias. A nonlinear strength of zero denotes correct specification of the corresponding working model.}
\label{fig:fixed-sensitivity-misspec-supp}
\end{figure}

We first vary the strengths of the nonlinear terms omitted from the working models for the propensity score and outcome regression around the main scenario \texttt{II} design, holding $n=1000$, the number of replications at 500, and the balancing basis from the main text fixed.
Specifically, the propensity score sensitivity parameter $\delta$ multiplies $|X_1X_2|$, and the sensitivity parameter $\lambda$ for the outcome regression multiplies $\mu_{\mathrm{nl}}$; each ranges over $\{0,0.5,1,1.5,2\}$, with $(\delta,\lambda)=(1,1)$ corresponding to the main simulation setting. Grid points with $\delta=0$ or $\lambda=0$ are correctly specified boundary cases for the corresponding working model.
Figure~\ref{fig:fixed-sensitivity-misspec-supp} shows that \texttt{Usual} and \texttt{CB-ols} become increasingly biased as the two forms of misspecification strengthen. At most grid points where both working models are misspecified, \texttt{CB-opt} has the smallest absolute bias, and its variation across the grid is smaller than that of the other estimators. Hence, the advantage of \texttt{CB-opt} in scenario \texttt{II} is not tied to the particular misspecification strengths used in the main simulation; it persists across a range of joint misspecification strengths.

\subsection{Sensitivity to the balancing basis}

\begin{figure}[htbp]
\centering
\includegraphics[width=\textwidth]{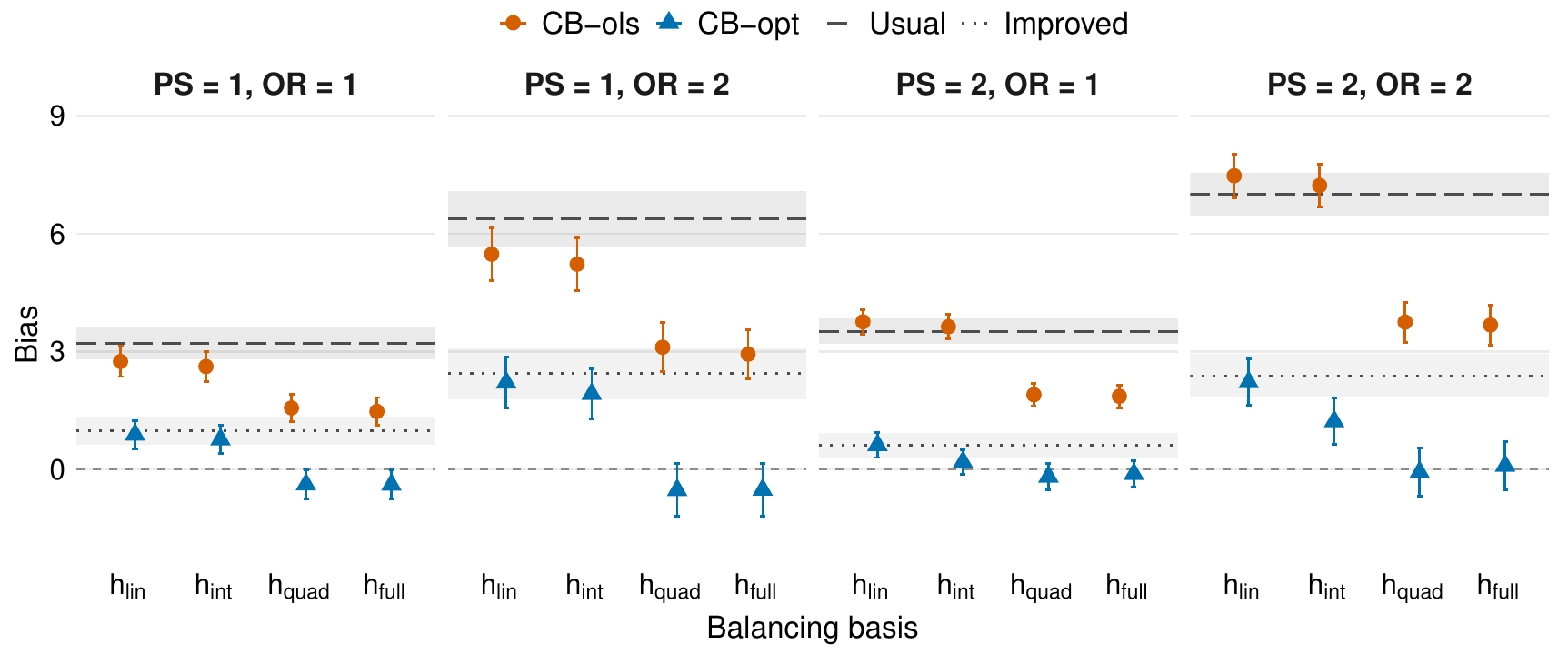}
\caption{Sensitivity of signed bias to the balancing basis in scenario \texttt{II} at $n=1000$. Points and error bars represent \texttt{CB-ols} and \texttt{CB-opt}; horizontal lines and bands represent \texttt{Usual} and \texttt{Improved}, with 95\% Monte Carlo intervals.}
\label{fig:fixed-sensitivity-basis-supp}
\end{figure}

We next compare four prespecified balancing bases:
$h_{\mathrm{lin}}=\{1,X_1,X_2\}$,
$h_{\mathrm{int}}=h_{\mathrm{lin}}\cup\{X_1X_2\}$,
$h_{\mathrm{quad}}=h_{\mathrm{lin}}\cup\{X_1^2,X_2^2\}$, and
$h_{\mathrm{full}}=h_{\mathrm{quad}}\cup\{X_1X_2\}$, where $h_{\mathrm{full}}$ is the basis used in the main text.
Figure~\ref{fig:fixed-sensitivity-basis-supp} reports signed bias at four combinations of the nonlinear strengths for the propensity score and outcome regression models. Across these configurations, \texttt{CB-opt} generally has smaller absolute bias than \texttt{Usual}, \texttt{Improved}, and \texttt{CB-ols}. Even with the linear basis $h_{\mathrm{lin}}$, its absolute bias is comparable to or smaller than that of \texttt{Improved} and is lower than that of \texttt{Usual} and \texttt{CB-ols}, showing that covariate balancing remains useful with a simple basis. The performance of \texttt{CB-opt} is similar under $h_{\mathrm{quad}}$ and $h_{\mathrm{full}}$, and both leave less residual bias than $h_{\mathrm{lin}}$ or $h_{\mathrm{int}}$. Thus, bias control depends on whether the balancing basis represents the main directions of misspecification. In this design, including the quadratic terms is more important than adding only $X_1X_2$. The relatively small basis $h_{\mathrm{full}}$ is therefore a practical default and supports the applicability of \texttt{CB-opt}.

The preceding comparison focuses on practical basis choice under joint misspecification. Supplementary Section~\ref{sec:app-or-route} shows that, when the propensity score model is misspecified, correct specification of the outcome regression model alone need not ensure consistency because the outcome parameter is selected by the variance criterion. It also gives a sufficient condition based on balancing the directions that enter the variance gradient. We denote the derivative-based balancing basis \(h(X;\widetilde\alpha,\widetilde\beta)\) in \eqref{eq:or-sufficient-balancing-functions} by \(h_{\mathrm{der}}\). We therefore compare the theory-guided basis $h_{\mathrm{der}}$ with the simpler polynomial basis $h_{\mathrm{full}}$, which is not constructed to match these directions exactly.

\begin{figure}[htbp]
\centering
\includegraphics[width=0.8\textwidth]{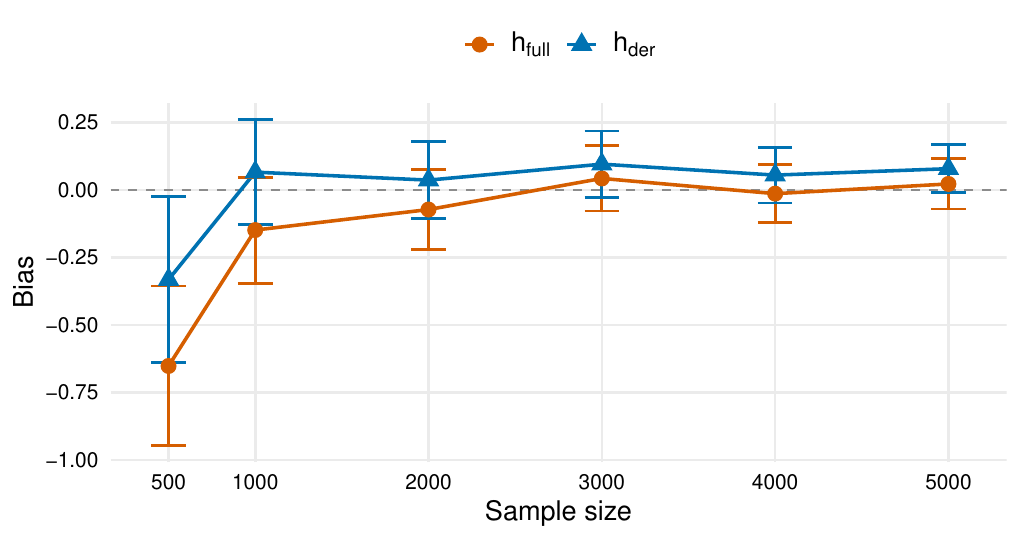}
\caption{Bias of the \texttt{CB-opt} estimator for rule $d_0$ in scenario \texttt{IC} across sample sizes, comparing the polynomial basis $h_{\mathrm{full}}$ with the theory-guided derivative basis $h_{\mathrm{der}}$ in Supplementary Section~\ref{sec:app-or-route}. Error bars are 95\% Monte Carlo intervals.}
\label{fig:fixed-derivative-ic-supp}
\end{figure}

Figure~\ref{fig:fixed-derivative-ic-supp} shows that $h_{\mathrm{der}}$ has smaller absolute bias at $n=500$ and $n=1000$, supporting the theoretical role of the derivative-based construction. From $n=3000$ onward, however, $h_{\mathrm{full}}$ has smaller absolute bias and remains close to zero. This pattern suggests that exact matching of the theoretical directions may not be needed for good finite-sample performance in this design. A simpler polynomial basis may approximate the important directions and may be less sensitive to numerical instability than the higher-dimensional $h_{\mathrm{der}}$. These results support $h_{\mathrm{full}}$ as a practical default in this design, while $h_{\mathrm{der}}$ illustrates the theory-guided basis construction motivated by the sufficient consistency condition in Supplementary Section~\ref{sec:app-or-route}.

\subsection{Implementation and average regret for learned rules}
\label{sec:app-learned-implementation}

For computation, we represent the distinct rules induced by \(\mathcal D\) on the simulation support as
\begin{equation}\label{eq:computational-rule-parameterization}
d_{b,\theta}(X)
=\ind\{b+\cos(\theta)X_1+\sin(\theta)X_2>0\},
~~|b|\leq2.5,\quad \theta\in[0,2\pi).
\end{equation}
To see the connection with the normalized representation in Section~\ref{sec:simulation}, consider any rule that is nonconstant on this support, for which $\eta_1\ne0$. Dividing its index by $\|\eta_1\|_2$, set $b=\eta_0/\|\eta_1\|_2$ and choose $\theta$ so that $\eta_1/\|\eta_1\|_2=(\cos\theta,\sin\theta)^\top$. This positive rescaling leaves the induced decision rule unchanged. Moreover, on the support \([-1,1]\times[-2,2]\),
the bound $\sup_X|\cos(\theta)X_1+\sin(\theta)X_2|\leq\sqrt5<2.5$ holds.
Consequently, values of $b$ beyond this range generate only the two constant rules, which are already represented by $b=2.5$ and $b=-2.5$. Thus, the bounded computational parameterization in \eqref{eq:computational-rule-parameterization} generates the same rules on the simulation support as the rule class in the main text while maintaining numerical stability.

Because the indicator function makes the empirical value criterion discontinuous and nonconvex in \((b,\theta)\), we use a multistart Nelder--Mead search implemented by the \texttt{optim()} function in R's \texttt{stats} package. Each search uses 20 random starting points together with a deterministic \(5\times7\) grid of starting values for $b$ and \(\theta\). Within each replication, the same generated data set and the same optimization starts are used across all four estimators. 

To evaluate the learned rules, we use the Cartesian product of 401 equally spaced cell midpoints in $[-1,1]$ for $X_1$ and 401 equally spaced cell midpoints in $[-2,2]$ for $X_2$, giving $401^2=160{,}801$ deterministic evaluation points.
We compute $V(d^{\opt})$ and the value of each selected rule by averaging with equal weights over this set.
The validation set is used only to evaluate a learned rule after training and is not used to fit nuisance models, tune the optimization, or select the rule.

\begin{figure}[htbp]
\centering
\includegraphics[width=\textwidth]{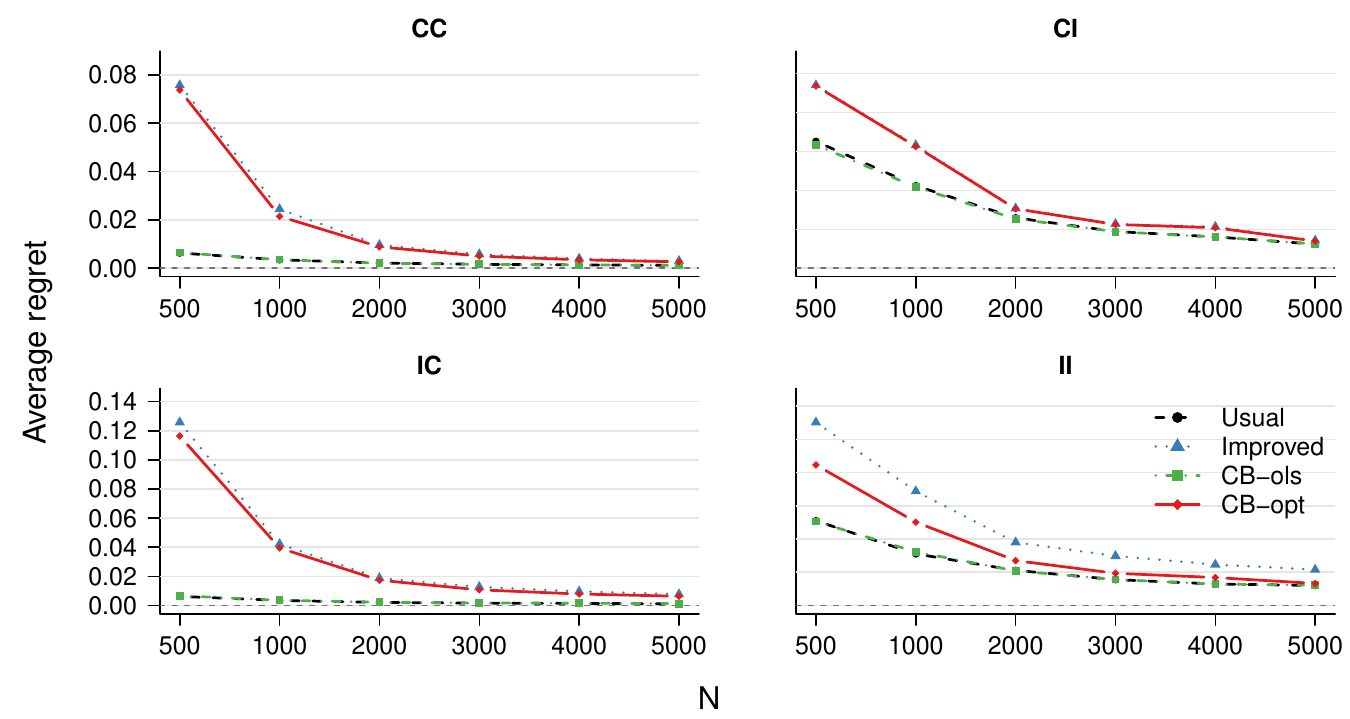}
\caption{Average regret of the learned rules based on $Z$ across the four nuisance model specification scenarios.}
\label{fig:learned-XZ-all-supp}
\end{figure}

Figure~\ref{fig:learned-XZ-all-supp} reports average regret for the analysis based on $Z$, using the same setting as Figure~\ref{fig:learned-X-all-main}. Regret decreases with sample size for every method. \texttt{Usual} and \texttt{CB-ols} generally have the smallest regret after adding $W$, with especially large gains in \texttt{CC} and \texttt{IC}; \texttt{Improved} and \texttt{CB-opt} benefit in those two scenarios but are less competitive in \texttt{CI} and \texttt{II}. Thus, auxiliary predictors do not uniformly improve rule learning even though they often improve efficiency for a given rule: because $W$ does not enlarge the rule class, it helps only when more precise value estimates change the ranking of candidate rules enough to offset the cost of estimating additional nuisance parameters.

Let $\overline R_X$ and $\overline R_Z$ denote the Monte Carlo mean regrets for the $X$- and $Z$-based analyses, respectively. Table~\ref{tab:learned-regret-representative-supp} reports these means, together with their Monte Carlo standard deviations, at $n=1000$ and $n=4000$. The last column gives $\overline R_Z-\overline R_X$, so a negative value favors the $Z$-based analysis, whereas a positive value favors the $X$-based analysis. In \texttt{CC}, all four methods achieve substantially lower regret with $Z$, and their performances become much closer. In \texttt{II}, adding $W$ yields only small improvements for \texttt{Usual} and \texttt{CB-ols} but increases regret for \texttt{Improved} and \texttt{CB-opt}; among the $X$-based procedures, \texttt{CB-opt} is comparatively favorable at the larger sample size. The table therefore qualifies the efficiency result in Proposition~\ref{prop:z-efficiency} for a fixed rule $d$: improved estimation of a prespecified value does not by itself guarantee better selection over a rule class, especially under joint misspecification.

\begin{table}[htbp]
\centering
\caption{Average regret at $n=1000$ and $n=4000$ for learned rules.}
\label{tab:learned-regret-representative-supp}
\begin{tabular}{cclccc}
\toprule
Scenario & $n$ & Method
& $X$-based $\overline R_X$ (MCSD)
& $Z$-based $\overline R_Z$ (MCSD)
& $\overline R_Z-\overline R_X$ \\
\midrule
\multirow{8}{*}{\texttt{CC}}
& \multirow{4}{*}{$1000$}
& \texttt{Usual}    & 0.3247 (0.3295) & 0.0035 (0.0038) & $-0.3212$ \\
&& \texttt{Improved} & 0.3749 (0.4027) & 0.0245 (0.0291) & $-0.3505$ \\
&& \texttt{CB-ols}   & 0.3232 (0.3255) & 0.0035 (0.0039) & $-0.3196$ \\
&& \texttt{CB-opt}   & 0.3749 (0.3880) & 0.0215 (0.0255) & $-0.3534$ \\
\cmidrule(lr){2-6}
& \multirow{4}{*}{$4000$}
& \texttt{Usual}    & 0.1366 (0.1234) & 0.0014 (0.0014) & $-0.1352$ \\
&& \texttt{Improved} & 0.1401 (0.1352) & 0.0040 (0.0044) & $-0.1361$ \\
&& \texttt{CB-ols}   & 0.1377 (0.1269) & 0.0014 (0.0014) & $-0.1363$ \\
&& \texttt{CB-opt}   & 0.1392 (0.1338) & 0.0035 (0.0038) & $-0.1357$ \\
\midrule
\multirow{8}{*}{\texttt{II}}
& \multirow{4}{*}{$1000$}
& \texttt{Usual}    & 0.3477 (0.3936) & 0.3104 (0.3562) & $-0.0373$ \\
&& \texttt{Improved} & 0.4795 (0.5858) & 0.6878 (1.0241) &  0.2083 \\
&& \texttt{CB-ols}   & 0.3384 (0.4016) & 0.3225 (0.3584) & $-0.0160$ \\
&& \texttt{CB-opt}   & 0.3503 (0.4174) & 0.5015 (0.6628) &  0.1512 \\
\cmidrule(lr){2-6}
& \multirow{4}{*}{$4000$}
& \texttt{Usual}    & 0.1390 (0.1354) & 0.1289 (0.1266) & $-0.0102$ \\
&& \texttt{Improved} & 0.1361 (0.1477) & 0.2459 (0.3221) &  0.1098 \\
&& \texttt{CB-ols}   & 0.1384 (0.1302) & 0.1285 (0.1247) & $-0.0099$ \\
&& \texttt{CB-opt}   & 0.1180 (0.1239) & 0.1679 (0.2191) &  0.0499 \\
\bottomrule
\end{tabular}
\end{table}

\subsection{Sensitivity to Repeated Random Splits in the Leukemia Application}
\label{sec:app-leukemia-split-sensitivity}

To assess sensitivity to the initial division into training and test samples, we repeated the random split 100 times, using 580 patients for training and the complementary 581 patients for testing in each repetition. For each method, the rule and nuisance models were estimated from the training sample, and the rule value was then estimated from the independent test sample after refitting the nuisance models. Table~\ref{tab:leukemia-split-sensitivity} summarizes the resulting value estimates.

This analysis describes sensitivity to the random split rather than providing an additional inferential procedure. Because the training and test samples overlap across repetitions, the resulting estimates are correlated. The SD reported in Table~\ref{tab:leukemia-split-sensitivity} denotes the standard deviation of the 100 value estimates across repeated splits and is not interpreted as an estimated standard error. The four distributions are broadly similar, indicating that the substantive scale of the value estimates is not driven by the single split used in the main analysis.

\begin{table}[htbp]
\centering
\caption{Sensitivity of the leukemia analysis to repeated random splits.}
\label{tab:leukemia-split-sensitivity}
\begin{tabular}{lrrrr}
\toprule
& \texttt{Usual} & \texttt{Improved} & \texttt{CB-ols} & \texttt{CB-opt} \\
\midrule
Mean & 1557 & 1574 & 1570 & 1569 \\
SD   &  102 &  114 &  100 &  112 \\
\bottomrule
\end{tabular}
\end{table}

\section{Proofs}
\label{sec:app-proofs}

\begin{proof}[Proof of Proposition~\ref{prop:double-robustness}]
For Proposition~\ref{prop:double-robustness}(i), correct specification of the propensity score model gives \(\kappa(X)=0\), and the result follows immediately from \eqref{eq:dr-value-decomp}.

For Proposition~\ref{prop:double-robustness}(ii), there exists a constant vector \(c\in\mathbb R^{\dim(\alpha)}\) such that
$q_d(X)=c^\top(G^*)^\top h(X;\alpha^*)$.
Therefore,
\[
\begin{aligned}
\E\left\{\kappa(X)q_d(X)\right\}
&=c^\top (G^*)^\top\E\left\{\kappa(X)h(X;\alpha^*)\right\}\\
&=c^\top (G^*)^\top\Psi(\alpha^*)=0
\end{aligned}
\]
by \eqref{eq:cb-population-foc}.
Thus \(\E\{\kappa(X)q_d(X)\}=0\), and the result again follows from \eqref{eq:dr-value-decomp}.
\end{proof}

\begin{proof}[Proof of Lemma~\ref{lem:cb-if}]
Under correct propensity score specification,
\[
\E\left\{\psi(A,X;\alpha^{\true})\mid X\right\}
=\frac{\E(A\mid X)-e_1(X;\alpha^{\true})}
{e_1(X;\alpha^{\true})\{1-e_1(X;\alpha^{\true})\}}
h(X;\alpha^{\true})=0.
\]
Uniqueness therefore implies $\alpha^*=\alpha^{\true}$.

Let $\ell_d(O;\alpha,\beta)$ denote the value estimating function in \eqref{eq:rule-value-function}.
At the true propensity score,
\[
\E\left\{ \frac{\partial\ell_d(O;\alpha^{\true},\beta)}{\partial\beta^\top} \right\} =-\E\left[ \frac{\ind\{A=d(X)\}-e_d^{\true}(X)}{e_d^{\true}(X)} \frac{\partial m_d(X;\beta)}{\partial\beta^\top} \right]=0.
\]
Thus estimating $\beta$ contributes only $o_p(1)$ to the first-order expansion, even when $\beta^*$ does not index the true outcome regression.
A Taylor expansion in $\alpha$ yields
\begin{align*}
\sqrt n\left\{\widehat V^{\cb}(d;\widehat\beta)-V(d)\right\} ={}&\frac{1}{\sqrt n}\sum_{i=1}^n \varphi^{\dr}(O_i;\alpha^{\true},\beta^*)\\
&+\E\left\{ \frac{\partial\varphi^{\dr} (O;\alpha^{\true},\beta^*)}{\partial\alpha^\top} \right\} \sqrt n(\widehat\alpha^{\cb}-\alpha^{\true})+o_p(1).
\end{align*}
Substituting \eqref{eq:alpha-if} gives \eqref{eq:value-if-expansion}.
\end{proof}

\begin{proof}[Proof of Proposition~\ref{prop:asymptotic-properties}]
Fix \(d\in\mathcal D\).
We first prove part~(i).
Under correct specification of the propensity score working model, \(\alpha^*=\alpha^{\true}\).
For any root-\(n\) consistent estimator
\(\widehat\beta\), Lemma~\ref{lem:cb-if} gives the expansion in \eqref{eq:value-if-expansion}.
Consequently, the first-order variance at the limiting outcome regression parameter \(\beta^*\) is $\mathcal S(\alpha^{\true},\beta^*)=\E[\{\varphi^{\cb}(O;\alpha^{\true},\beta^*)\}^2]$.
By definition, \(\beta^{\opt}\in\argmin_\beta\mathcal S(\alpha^{\true},\beta)\), and consistency of the empirical minimizer gives \(\widehat\beta^{\opt}\overset{p}{\to}\beta^{\opt}\).
Thus \(\widehat V^{\cb}(d)\) attains the smallest first-order variance among the estimators in \eqref{eq:cb-value-class}, proving part~(i).

We next prove part~(ii) using standard \(M\)-estimation theory.
Let \((\alpha^\dagger,\beta^\dagger)\) denote the population nuisance target under the applicable condition in Proposition~\ref{prop:double-robustness}.
Thus \((\alpha^\dagger,\beta^\dagger)=(\alpha^{\true},\beta^{\opt})\) under correct propensity score specification, whereas \((\alpha^\dagger,\beta^\dagger)=(\alpha^*,\beta^*)\) under condition~\eqref{eq:qd-effective-space}.
In either case, Proposition~\ref{prop:double-robustness} gives the equation
\[
\E\left\{\ell_d(O;\alpha^\dagger,\beta^\dagger)\right\}=V(d),
\]
where $\ell_d(O;\alpha,\beta)$ is the value estimating function defined in \eqref{eq:rule-value-function}.

We now write the estimating equations for the nuisance parameters explicitly. Introduce auxiliary parameters \(\Psi\), \(G\), \(K\), \(\Gamma\), \(D\), and \(\Upsilon\). Here \(\Psi\), without an argument, denotes the auxiliary vector whose population target is \(\Psi(\alpha)=\E\{\psi(A,X;\alpha)\}\), and the target of \(G\) is \(G(\alpha)=\E\{\partial\psi(A,X;\alpha)/\partial\alpha^\top\}\). Set \(K=(G^\top G)^{-1}G^\top\). At \(\alpha=\alpha^{\true}\), this coincides with the matrix \(K\) defined after \eqref{eq:cb-estimating-equation}; under the setup of Supplementary Section~\ref{sec:app-or-route}, its value at \(\alpha^*\) is \(K^*\). Finally, let \(\Gamma=-\E\{\partial\ell_d(O;\alpha,\beta)/\partial\alpha^\top\}\), \(D=-\E\{\partial^2\ell_d(O;\alpha,\beta)/(\partial\beta\,\partial\alpha^\top)\}K\), and
\[
\Upsilon
=\E\left\{
\frac{\partial\ell_d(O;\alpha,\beta)}{\partial\beta}
+D\psi(A,X;\alpha)
\right\}.
\]
Thus, \(D\) accounts for the dependence of \(\Gamma K\) on \(\beta\), while \(\Upsilon\) centers the corresponding derivative in the first-order condition for variance minimization. Under the setup of Supplementary Section~\ref{sec:app-or-route}, the value of \(D\) at \((\alpha^*,\beta^{\true})\) is \(D^*\). Let \(\xi\) collect \(\alpha\), \(\beta\), \(\Psi\), \(G\), \(K\), \(\Gamma\), \(D\), and \(\Upsilon\), and let \(\operatorname{vec}(M)\) denote the vector obtained by stacking the columns of a matrix \(M\).
For the identity-weighted balancing criterion used in \eqref{eq:cb-estimating-equation}, the stacked estimating function is
\begin{equation}
U(O;\xi,V)
=
\begin{pmatrix}
G^\top\Psi\\
\psi(A,X;\alpha)-\Psi\\
\operatorname{vec}\left\{\dfrac{\partial\psi(A,X;\alpha)}{\partial\alpha^\top}-G\right\}\\
\operatorname{vec}\{K-(G^\top G)^{-1}G^\top\}\\
\Gamma^\top+\dfrac{\partial\ell_d(O;\alpha,\beta)}{\partial\alpha}\\
\operatorname{vec}\left\{D+
\dfrac{\partial^2\ell_d(O;\alpha,\beta)}
{\partial\beta\,\partial\alpha^\top}K\right\}\\
\dfrac{\partial\ell_d(O;\alpha,\beta)}{\partial\beta}
+D\psi(A,X;\alpha)-\Upsilon\\
2\left\{\ell_d(O;\alpha,\beta)-V
+\Gamma K\psi(A,X;\alpha)\right\}
\left\{\dfrac{\partial\ell_d(O;\alpha,\beta)}{\partial\beta}
+D\psi(A,X;\alpha)-\Upsilon\right\}
\end{pmatrix}.
\end{equation}
The first three blocks imply \(G^\top\Psi=0\), \(\Psi=\Psi(\alpha)=\E\{\psi(A,X;\alpha)\}\), and \(G=G(\alpha)=\E\{\partial\psi(A,X;\alpha)/\partial\alpha^\top\}\), so their sample analogues reproduce the first-order condition for minimizing \(\widehat Q_n(\alpha)\). The next three blocks estimate \(K\), \(\Gamma\), and the derivative of \(\Gamma K\) with respect to \(\beta\), and the following block estimates \(\Upsilon\). The first four blocks give \(K\Psi=0\), and the value equation below gives \(\Pn\{\ell_d(O;\alpha,\beta)-V\}=0\). Hence the empirical mean of the first factor in the last block is zero, while the estimating equation for \(\Upsilon\) centers its second factor. The last block therefore gives the first-order condition for minimizing \(\widehat{\mathcal S}_n\) with respect to \(\beta\).
Write \(\xi^\dagger\) for its population solution, and define \(\zeta=\{\xi^\top,V\}^\top\) and \(\zeta^\dagger=\{(\xi^\dagger)^\top,V(d)\}^\top\).
The stacked estimating function for the nuisance parameters and
value is
\begin{equation}
m^{\opt}(O;\zeta)
=
\begin{pmatrix}
U(O;\xi,V)\\
\ell_d(O;\alpha,\beta)-V
\end{pmatrix}.
\end{equation}
Let \(\widehat\zeta=\{(\widehat\xi)^\top,\widehat V^{\cb}(d)\}^\top\) denote the joint sample estimator used to construct the plug-in sandwich variance estimator described in Section~\ref{sec:cb-asymptotics}. It satisfies $\Pn m^{\opt}(O;\widehat\zeta)=o_p(n^{-1/2})$.
Standard \(M\)-estimation theory yields
\begin{equation}\label{eq:stacked-m-asymptotic-normality}
\sqrt n(\widehat\zeta-\zeta^\dagger)
\overset{d}{\longrightarrow}
N\!\left(
0,\,
B(\zeta^\dagger)^{-1}C(\zeta^\dagger)
\{B(\zeta^\dagger)^{-1}\}^{\top}
\right),
\end{equation}
where
\begin{equation}
B(\zeta)
=\E\left\{\frac{\partial m^{\opt}(O;\zeta)}
{\partial\zeta^\top}\right\},
~~
C(\zeta)
=\E\left\{m^{\opt}(O;\zeta)m^{\opt}(O;\zeta)^\top\right\}.
\end{equation}
Partition these matrices at \(\zeta^\dagger\) according to the nuisance block \(\xi\) and the scalar value \(V\). Only the last block of $U(O;\xi,V)$ depends on $V$, and the estimating equation defining \(\Upsilon\) gives
\[
-2\E\left\{
\frac{\partial\ell_d(O;\alpha,\beta)}{\partial\beta}
+D\psi(A,X;\alpha)-\Upsilon
\right\}=0.
\]
The upper-right block is therefore zero, and
\begin{equation}\label{eq:stacked-bc-partition}
B(\zeta^\dagger)
=
\begin{pmatrix}
B_1&0\\
B_2&-1
\end{pmatrix},
~~
C(\zeta^\dagger)
=
\begin{pmatrix}
C_{11}&C_{12}\\
C_{12}^\top&C_{22}
\end{pmatrix}.
\end{equation}
Taking the value component of \eqref{eq:stacked-m-asymptotic-normality} and using the partition in \eqref{eq:stacked-bc-partition} gives
\begin{equation}
\sqrt n\left\{\widehat V^{\cb}(d)-V(d)\right\}
\overset{d}{\longrightarrow}N(0,\Lambda),
\end{equation}
where $\Lambda=B_2B_1^{-1}C_{11}(B_1^\top)^{-1}B_2^\top-2B_2B_1^{-1}C_{12}+C_{22}.$
Replacing these population blocks by their empirical analogues evaluated at \(\widehat\zeta\) gives \(\widehat\Lambda\overset{p}{\to}\Lambda\) under the same uniform convergence and nonsingularity conditions.
This proves part~(ii).

Finally, suppose both nuisance working models are correctly specified. Then \(\beta^{\true}\) belongs to the outcome parameter space and provides a feasible benchmark for the variance minimization problem. At \((\alpha^{\true},\beta^{\true})\), conditioning on \(X\) gives $\Gamma(\alpha^{\true},\beta^{\true})=-\E\{\partial\varphi^{\dr}(O;\alpha^{\true},\beta^{\true})/\partial\alpha^\top\}=0$, so the correction due to propensity score estimation in \eqref{eq:cb-value-if} vanishes. Substituting \(m_d(X;\beta^{\true})=m_d^{\true}(X)\) into \eqref{eq:dr-if} gives
\begin{equation}\label{eq:efficient-if}
\varphi^{\cb}(O;\alpha^{\true},\beta^{\true})
=\varphi^{\dr}(O;\alpha^{\true},\beta^{\true})
:=m_d^{\true}(X)
+\frac{\ind\{A=d(X)\}}{e_d^{\true}(X)}
\{Y-m_d^{\true}(X)\}
-V(d).
\end{equation}
The function \(\varphi^{\dr}(O;\alpha^{\true},\beta^{\true})\) in \eqref{eq:efficient-if} is the efficient influence function for the value of a fixed rule in the nonparametric model under Assumption~\ref{assump:regular}.
Conditioning on \(X\), its variance is
\begin{equation}\label{eq:semiparametric-eff-bound}
\begin{aligned}
\operatorname{var}\{\varphi^{\dr}(O;\alpha^{\true},\beta^{\true})\}
={}&\E\!\left[
\frac{d(X)\operatorname{var}(Y\mid A=1,X)}{e_1^{\true}(X)}
+\frac{\{1-d(X)\}\operatorname{var}(Y\mid A=0,X)}{e_0^{\true}(X)}
\right]\\
&+\E\!\left[\{m_d^{\true}(X)-V(d)\}^2\right],
\end{aligned}
\end{equation}
The quantity in \eqref{eq:semiparametric-eff-bound} is the semiparametric efficiency bound. To compare it with the criterion at an arbitrary \(\beta\), define
\[
\Delta_\beta(O)=\varphi^{\cb}(O;\alpha^{\true},\beta)
-\varphi^{\dr}(O;\alpha^{\true},\beta^{\true}).
\]
Under joint correct specification, \(\Delta_\beta(O)\) is a function of \((A,X)\) and \(\E\{\Delta_\beta(O)\mid X\}=0\). Because \(\E\{Y-m_d^{\true}(X)\mid A=d(X),X\}=0\), iterated expectations give
\[
\begin{aligned}
\E\{\varphi^{\dr}(O;\alpha^{\true},\beta^{\true})\Delta_\beta(O)\}
&=\E\left[\{m_d^{\true}(X)-V(d)\}\Delta_\beta(O)\right]\\
&=\E\left[\{m_d^{\true}(X)-V(d)\}\E\{\Delta_\beta(O)\mid X\}\right]=0.
\end{aligned}
\]
Consequently,
\[
\mathcal S(\alpha^{\true},\beta)
=\operatorname{var}\{\varphi^{\dr}(O;\alpha^{\true},\beta^{\true})\}
+\E\left\{\Delta_\beta(O)^2\right\}
\geq\operatorname{var}\{\varphi^{\dr}(O;\alpha^{\true},\beta^{\true})\}.
\]
Because \(\beta^{\true}\) is feasible and \(\Delta_{\beta^{\true}}=0\), minimizing over \(\beta\) gives \(\mathcal S(\alpha^{\true},\beta^{\opt})=\operatorname{var}\{\varphi^{\dr}(O;\alpha^{\true},\beta^{\true})\}\). Therefore, \(\Lambda\) equals the semiparametric efficiency bound, and \(\widehat V^{\cb}(d)\) attains this bound, proving part~(iii).
\end{proof}

\begin{proof}[Proof of Proposition~\ref{prop:z-efficiency}]
Fix an arbitrary $\beta_{\subX}$.
By Assumption~\ref{assump:nested-first-order}(ii), choose $\beta_{\subZ}$ such that the fitted outcome models agree for both treatment levels.
Together with correct propensity score specification, this gives
\begin{equation*}
\varphi_{\subZ}^{\dr} \left(O;\alpha^{\true},\beta_{\subZ}\right) = \varphi_{\subX}^{\dr}(O;\alpha^{\true},\beta_{\subX}) \quad\text{almost surely}.
\end{equation*}
Moreover, Assumption~\ref{assump:nested-first-order}(i) and (iii) imply that both analyses use the same estimated propensity score. Thus, at these paired outcome-model parameter values, the two value estimators agree, and so do their first-order representations:
\begin{equation*}
\varphi_{\subZ}^{\cb} \left(O;\alpha^{\true},\beta_{\subZ}\right) = \varphi_{\subX}^{\cb}(O;\alpha^{\true},\beta_{\subX}) \quad\text{almost surely}.
\end{equation*}
It follows that $\mathcal S_{\subZ}(\beta_{\subZ})=\mathcal S_{\subX}(\beta_{\subX})$ for every such pair.
In particular, Assumption~\ref{assump:nested-first-order}(ii) ensures that there exists a $Z$-based parameter value $\widetilde\beta_{\subZ}$ such that
\[
m_a(Z;\widetilde\beta_{\subZ})
=m_a(X;\beta_{\subX}^{\opt})
\quad\text{almost surely for }a\in\{0,1\}.
\]
Then
\[
\mathcal S_{\subZ}(\beta_{\subZ}^{\opt})
\leq \mathcal S_{\subZ}(\widetilde\beta_{\subZ})
=\mathcal S_{\subX}(\beta_{\subX}^{\opt}).
\]
This proves \eqref{eq:variance-ordering}.
\end{proof}

\begin{proof}[Proof of Proposition~\ref{prop:or-route-sufficient}]
Suppress the arguments $(X;\alpha^*,\beta^{\true})$. Correct outcome regression specification implies $\E\{Y-m_d^{\true}(X)\mid A=d(X),X\}=0$, and hence $\Gamma(\alpha^*,\beta^{\true})=\E[\ind\{A=d(X)\}\{Y-m_d^{\true}(X)\}e_{d,\alpha}^\top/e_d^2]=0$. Consequently, $\varphi^{\cb}=\varphi^{\dr}$ at $\beta^{\true}$, and
\[
\left.\frac{\partial\varphi^{\cb}(O;\alpha^*,\beta)}{\partial\beta}\right|_{\beta=\beta^{\true}}
=\left[1-\frac{\ind\{A=d(X)\}}{e_d}\right]m_{d,\beta}+D^*\psi(A,X;\alpha^*).
\]
Differentiating the second-moment criterion therefore gives
\[
\nabla_\beta\mathcal S(\alpha^*,\beta^{\true})
=2\E\left[\varphi^{\dr}\left\{\left[1-\frac{\ind\{A=d(X)\}}{e_d}\right]m_{d,\beta}+D^*\psi(A,X;\alpha^*)\right\}\right].
\]

We next evaluate the two terms in this expectation conditional on $X$. For the first term, the contribution involving $Y-m_d^{\true}(X)$ is zero by correct outcome regression specification. Moreover, $\E[1-\ind\{A=d(X)\}/e_d\mid X]=1-e_d^{\true}/e_d$. Using $e_d^{\true}-e_d=\{2d(X)-1\}\{e_1^{\true}(X)-e_1(X;\alpha^*)\}$, $e_d(1-e_d)=e_1(X;\alpha^*)\{1-e_1(X;\alpha^*)\}$, and the definition of $\kappa(X)$, we obtain
\[
\E\left\{\left.\varphi^{\dr}\left[1-\frac{\ind\{A=d(X)\}}{e_d}\right]m_{d,\beta}\right|X\right\}
=-\kappa(X)\{2d(X)-1\}(1-e_d)\{m_d^{\true}(X)-V(d)\}m_{d,\beta}.
\]
For the second term, $\psi(A,X;\alpha^*)=\{A-e_1(X;\alpha^*)\}h^*(X)/[e_1(X;\alpha^*)\{1-e_1(X;\alpha^*)\}]$, so $\E\{\psi(A,X;\alpha^*)\mid X\}=\kappa(X)h^*(X)$. The residual contribution is again zero, yielding
\[
\E\{\varphi^{\dr}\psi(A,X;\alpha^*)\mid X\}
=\kappa(X)\{m_d^{\true}(X)-V(d)\}h^*(X).
\]
Combining these two identities with the gradient expression gives \eqref{eq:or-variance-gradient-representation}, because the remaining factor is precisely $b_d(X)$ defined in \eqref{eq:or-gradient-direction}.

By \eqref{eq:cb-population-foc}, $\E\{\kappa(X)f(X)\}=0$ for every $f\in\mathcal H_{\rm eff}$. The two maintained componentwise conditions therefore imply $\E\{\kappa(X)b_d(X)\}=0$ and $\E[\kappa(X)m_d(X;\beta^{\true})b_d(X)]=0$. Expanding the right-hand side of \eqref{eq:or-variance-gradient-representation} now gives $\nabla_\beta\mathcal S(\alpha^*,\beta^{\true})=0$. Strict convexity implies $\beta^*=\beta^{\true}$. Hence $q_d(X)=0$, and Proposition~\ref{prop:double-robustness}(ii) gives $\widehat V^{\cb}(d)\overset{p}{\to}V(d)$.
\end{proof}

\begin{proof}[Proof of Theorem~\ref{thm:regret-rate}]
Define $\mathcal E_n=\sup_{d\in\mathcal D_n}|\widehat V^{\cb}(d)-V(d)|$.
For every $d\in\mathcal D_n$, add and subtract $\Pn\ell_d^*(O)$ and $\E\ell_d^*(O)$ to obtain
\begin{align*}
\widehat V^{\cb}(d)-V(d) =\Pn\left\{\ell_d(O;\widehat\alpha^{\cb},\widehat\beta_d^{\opt}) -\ell_d^*(O)\right\}+(\Pn-\Pp)\ell_d^*(O)+\E\left\{\ell_d^*(O)\right\}-V(d).
\end{align*}
By Assumption~\ref{assump:uniform-rule-learning}, these three terms are uniformly $O_p(r_{\alpha,n}+r_{\beta,n})$, $O_p\{(v_n/n)^{1/2}\}$, and at most $b_n$, respectively.
Therefore,
\begin{equation}\label{eq:uniform-value-rate}
\mathcal E_n =O_p\left( \sqrt{\frac{v_n}{n}} +r_{\alpha,n}+r_{\beta,n}+b_n \right).
\end{equation}

Combining \eqref{eq:uniform-value-rate} with approximate empirical maximization gives $V(d_n^{\opt})-V(\widehat d_n) \leq\{V(d_n^{\opt})-\widehat V^{\cb}(d_n^{\opt})\} +\{\widehat V^{\cb}(\widehat d_n)-V(\widehat d_n)\}+\rho_n \leq2\mathcal E_n+\rho_n$, which proves the result.
\end{proof}

\end{document}